\documentclass[12pt,preprint]{aastex}
\usepackage{longtable}

\shorttitle{Monte Carlo Cluster Evolution V.}
\shortauthors{Chatterjee, Fregeau, Umbreit, \& Rasio}

\begin{document}
\title{Monte Carlo Simulations of Globular Cluster Evolution. V. Binary Stellar Evolution}
\author{Sourav Chatterjee\altaffilmark{1}, John M.\ Fregeau\altaffilmark{2,3}, Stefan Umbreit\altaffilmark{1}, and Frederic A.\ Rasio\altaffilmark{1}}
\affil{$^{1}$ Department of Physics and Astronomy, Northwestern University, 
Evanston, IL 60208, USA}
\affil{$^{2}$ Kavli Institute for Theoretical Physics, UCSB, Santa Barbara, CA 93106, USA}
\affil{$^{3}$ Chandra/Einstein Fellow}

\begin{abstract}
\label{abstract}
We study the dynamical evolution of globular clusters containing primordial binaries, including full 
single and binary stellar evolution using our Monte Carlo cluster evolution code updated with 
an adaptation of the single and binary stellar evolution codes SSE/BSE 
from \citet{2000MNRAS.315..543H,2002MNRAS.329..897H}.  We describe the 
modifications we have made to the code.  We present several test calculations and comparisons 
with existing studies to illustrate the validity of the code.  We show that our code finds very good 
agreement with direct {\em N}-body simulations including primordial binaries and stellar 
evolution.  We find significant 
differences in the evolution of the global properties of the simulated clusters using stellar evolution 
compared to simulations without any stellar evolution.  In particular, we find that the 
mass loss from stellar evolution acts as a significant energy production channel simply by reducing 
the total gravitational binding energy and can significantly prolong the initial core contraction 
phase before reaching the binary-burning quasi steady state of the cluster evolution as noticed in 
Paper IV.  We simulate a large grid of clusters varying the initial cluster mass, binary fraction, 
and concentration and compare properties of the simulated clusters with those of the observed 
Galactic globular clusters (GGCs).  We find that our simulated cluster properties agree well with the observed GGC properties.  We explore in some detail qualitatively different clusters in 
different phases of their evolution, and construct synthetic Hertzprung-Russell 
diagrams for these clusters.  
\end{abstract}
\section{Introduction}
\label{intro}

Star clusters in general, and the Galactic globular clusters in particular, have been studied extensively for many years.  As tracers
of the Galaxy's potential, their dynamical history tells us something about the formation and evolution of our Galaxy.
As dense stellar environments, their interesting constituent populations (including, e.g., blue straggler stars, cataclysmic
variables, and low-mass X-ray binaries) inform our understanding of binary stellar evolution through its interaction
with dynamics.  The study of the evolution of globular and other dense star clusters has had a somewhat long and varied 
history \citep[e.g.,][]{2003gmbp.book.....H}.  Before observations showed that globular clusters contained dynamically significant 
numbers of binaries, theoretical efforts focused on understanding the process of core collapse and the ensuing post-collapse evolution 
driven by three-body binary formation.  Once it became clear in the early 90s from observations that clusters contained sufficient numbers of binaries
that they must have been born with substantial ``primordial'' populations, theory emphasized properties
of clusters in the ``binary burning'' phase, in which the cluster core is supported against collapse by
super-elastic dynamical scattering interactions of binary stars.  More recently it has been realized
that pure point-mass interactions of binaries result in equilibrium cluster core radii in the binary
burning phase that are a factor of $\sim 10$ smaller than what is observed, and so many efforts have focused
on alternative cluster energy sources such as central intermediate-mass black holes, expedited
stellar mass loss from compact object formation via collisions, or prolonged mass segregation of compact objects 
\citep{2006MNRAS.368..677H,2007ApJ...658.1047F,2006astro.ph.12040T,2008IAUS..246..151C,2004ApJ...608L..25M}.
In a similar vein, recent theoretical work, combined with observations
showing that the core binary fraction in many clusters is fairly low ($\lesssim 10\%$), suggest 
that clusters may be born with remarkably low binary fractions of just a few percent \citep{2009arXiv0907.4196F}.
Such a small primordial binary fraction would be surprising since observations of young stars suggest that star formation
yields binary fractions of order $\sim 50\%$.  Clearly, our understanding of globular cluster evolution has changed
considerably over the past few decades, and much of that change has been driven by numerical simulations.

Among computational tools for studying the dynamical evolution of star clusters, the H\'enon Monte Carlo (MC) technique
\citep{1971Ap&SS..13..284H,1971Ap&SS..14..151H}
represents a balanced compromise between exactness and speed.  The MC method allows for a star-by-star realization
of the cluster, with its $N$ mass shells representing the $N$ stellar objects in the cluster (either single
or binary stars).  It assumes, most importantly, spherical symmetry and diffusive two-body relaxation, allowing time
integration on a relaxation timescale, and 
a computational cost that scales as $N \log N$.  We have developed our H\'enon MC cluster evolution code 
(which we call CMC, for ``Cluster Monte Carlo'') over the past decade 
\citep[][henceforth Papers I, II, III, and IV, respectively]{2000ApJ...540..969J,2001ApJ...550..691J,2003ApJ...593..772F,2007ApJ...658.1047F}.  
Since it allows for a star-by-star description of the cluster at each
timestep, it is relatively easy to add physical processes beyond two-body relaxation to the code.
We have previously added the effects of a Galactic tidal field, dynamical scattering interactions of binary star systems, and physical collisions
between stars.  In this paper we describe the addition of stellar evolution of single and binary star systems.
Many stars in a cluster evolve internally on a timescale shorter than the age of the cluster.
At early times they may lose a substantial fraction of their mass via stellar winds.  At later
times they may evolve off the main sequence, changing their masses and radii 
(and hence collision cross section), and
possibly receiving systemic velocity kicks when they become compact objects.  Since the binding energy of binary stars
is an important fuel source that can postpone the deep core collapse of star clusters, stellar evolution of binary
systems directly affects their global evolution.  Conversely, the properties of the cluster environment feed back on 
stellar evolution, modifying the evolutionary pathways of binary systems and the properties and numbers of interesting
binary systems relative to the low-density Galactic field 
\citep[e.g., X-ray binaries;][]{2006MNRAS.372.1043I,2008MNRAS.386..553I}.

Previous cluster evolution studies that include stellar evolution have improved our understanding of the global
evolution of clusters greatly, identifying several distinct stages of evolution. At early times, as the stars are
forming and the most massive stars have already begun nuclear burning, the cluster loses mass through residual gas
expulsion and stellar winds, resulting in cluster expansion during the first few Myr of evolution 
\citep{2001MNRAS.323..630H,2002MNRAS.329..897H}.  Shortly thereafter, if a runaway collision
scenario is avoided \citep[e.g.,][]{2006MNRAS.368..141F}, two-body relaxation dominates, 
resulting in a fairly long-lived (from a few to tens of Gyr) phase of core contraction.  
Once the core density becomes high enough for the energy generated in binary scattering interactions to balance the energy
carried out of the core by two-body relaxation, a potentially {\em very} long-lived (up to tens of Hubble times or more) 
phase of ``binary burning'' ensues \citep[e.g.,][]{2007MNRAS.379...93H,2007ApJ...658.1047F}.  
Once the population of binaries is exhausted in the core, the cluster goes into deep core collapse. In
the classical, point-mass limit, deep core collapse is arrested by the formation of a ``three-body binary'' 
and followed by a phase of gravothermal oscillations \citep{2003gmbp.book.....H}.  However, three-body 
binary formation may be inhibited by stellar collisions in sufficiently young and massive
clusters \citep{2006MNRAS.368..121F}.

With the exception of a few recent simulations, most numerical studies that include 
stellar evolution have either been limited in the number of stars they can treat
or have adopted a narrow initial mass function with very simplified stellar evolution recipes 
\citep[e.g.,][]{1998MNRAS.298.1239G,2000MNRAS.317..581G,2000ApJ...540..969J,2001ApJ...550..691J,2003ApJ...593..772F,2007ApJ...658.1047F}.  
Stars in star clusters are born with a range of masses up to $\sim 100 \, M_\sun$, and down to at least the hydrogen-burning
limit \citep[e.g.,][]{1955ApJ...121..161S,1979ApJS...41..513M,2001MNRAS.322..231K}, so one should evolve 
the full spectrum of stellar masses as realistically as possible to properly treat the influence
of stellar evolution on global cluster evolution.  Emphasis has recently been placed on comparing
observed properties of Galactic globular clusters with theoretical predictions.
Comparison of observed cluster structural properties with theory \citep[e.g.,][]{2007MNRAS.379...93H,2007ApJ...658.1047F}
suggests that either an additional energy source is ``puffing'' up cluster cores 
\citep{2008MNRAS.tmp..374M,2007MNRAS.374..857T,2008IAUS..246..151C}, or perhaps the clusters
are not in the expected evolutionary states, namely binary-burning \citep{2008ApJ...673L..25F}.
For example, recent $N$-body simulations by \citet{2007MNRAS.379...93H} show that the core
contraction phase can last a Hubble time, resulting in a cluster core radius that is larger
than one would expect were the cluster in the binary burning phase.  In these simulations,
the core contraction phase is prolonged by mass loss from stellar evolution.
Clearly, stellar evolution may be an important component in globular cluster evolution.

To more properly treat stellar evolution, we have recently coupled to CMC the stellar
evolution recipes of \citet[][hereafter referred to as BSE]{2000MNRAS.315..543H,2002MNRAS.329..897H}.
We choose BSE for ease of implementation, and for more direct comparisons with $N$-body calculations,
which commonly use BSE for stellar evolution.  In \S\ref{method} we describe the implementation
of stellar evolution in our code.  In \S\ref{comparison} we validate it by comparing with existing cluster evolution
calculations in the literature.  In \S\ref{without} we demonstrate the importance of stellar evolution
by comparing simulations that do not include it.  In \S\ref{real} we apply our newly updated code to the evolution
of a large grid of cluster models, highlighting typical behavior and comparing with observations.
Finally, in \S\ref{conclusion} we summarize and conclude.

\section{Method}
\label{method}
CMC treats a number of important physical processes, including 
two-body relaxation, the tidal effects of a host galaxy, strong binary-binary (BB) and binary-single (BS) 
scattering interactions, and direct physical collisions between stars (Paper IV).  
Here we describe the recent addition of BSE \citep{2000MNRAS.315..543H,2002MNRAS.329..897H} 
to treat stellar evolution of single and binary stars.

\subsection{Stellar Evolution}
\label{stellar_evolution}
For ease of implementation and for more direct comparisons with direct $N$-body we use the BSE stellar
evolution routines, as described in detail in \citet{2000MNRAS.315..543H,2002MNRAS.329..897H}.
For single stars, BSE comprises analytic functional fits to theoretically calculated stellar evolution
tracks as a function of metallicity and mass.  Binary star systems are treated using the same
fitting formulae for each star, but with treatments of physics relevant to binaries, including
stable and unstable mass transfer, common envelope evolution, magnetic braking, tidal coupling, and
the effects of neutron star and black hole birth kicks.  As described above, CMC allows for a 
star-by-star description of the cluster at every timestep, allowing for the inclusion of 
additional physics.  The stellar properties of binary and single stars are updated in step with
dynamics via function calls to the BSE library.  As described in detail in Papers I--IV, 
CMC uses a shared timestep which must be set as small as the smallest characteristic timescale
for each physical process.  We set the characteristic timescale for stellar evolution
to the time for the cluster to lose a fraction $0.001$ of its total mass.  In this way, any cluster
expansion from stellar evolutionary mass loss is properly resolved.

One aspect of our BSE implementation requires special mention, however.  
The evolution of high mass 
stars ($\gtrsim 100 \, M_\sun$) is rather uncertain and can vary greatly depending on the wind mass loss prescription.
These high mass stars are also quite rare and short lived, so observational constraints are limited.  
In BSE stars with mass $>100\,M_\sun$ are evolved as if they were $100 \, M_\sun$ stars.  When their 
dynamical properties are returned from BSE, their original ($> 100 \, M_\sun$) mass is returned.

\subsection{Collisions}
\label{collision}
As described in Paper IV,
collisions are treated in the sticky-sphere approximation, which was shown to be remarkably
accurate for the velocity dispersions found in globular clusters \citep{2006MNRAS.368..141F,2006MNRAS.368..121F}.  
When two stars collide, their properties (e.g., stellar type and effective age) are
set according to the BSE merger matrix and prescriptions as described in \citet{2002MNRAS.329..897H}.
In CMC this is implemented in the following simple way.  The two stars are passed to BSE as a very 
tight, eccentric binary with nearly zero pericenter distance and evolved for a very short time until 
they merge.  The properties of the merger product are then naturally set within BSE, and returned
to CMC as a single star.  BSE by default assumes full mixing of nuclear fuel during a collision involving 
MSSs.  We adopt this same rejuventation prescription for the simulations presented in this paper, but
note that the amount of mixing in the collision product should depend on the details of the interaction 
parameters leading to the collision, as well as the evolutionary stages of the collision 
progenitors \citep[e.g., ][]{1995ApJ...445L.117L,1996ApJ...468..797L,1997ApJ...487..290S,2001ApJ...548..323S,2002ApJ...568..939L}.  
In fact, it is found using detailed SPH calculations that the amount of mixing as a result of a collision may 
be minimal, especially for collisions involving evolved stars \citep{1995ApJ...445L.117L,1996ApJ...468..797L}.  

\subsection{Tidal Truncation Treatment}
\label{tide}
\begin{figure}
\begin{center}
\plotone{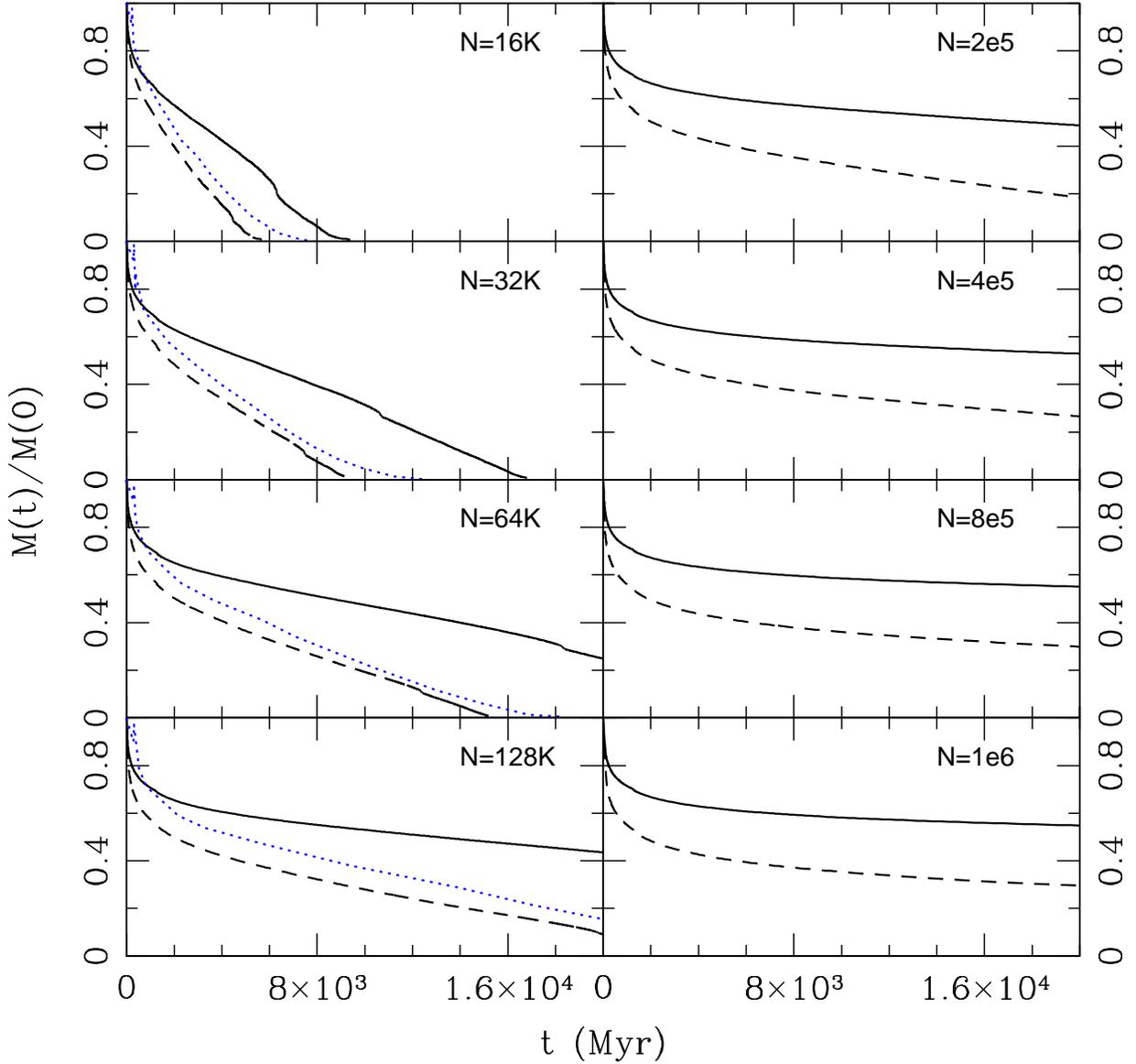}
\caption{Evolution of the total cluster mass for clusters with various initial 
numbers of stars and masses.  In each case a Galactocentric distance $r_{GC} = 8.5\,\rm{kpc}$ 
and a standard Galactic tidal field are assumed.  The initial number of stars 
($N$) is noted in each panel.  
The solid black line and the dashed black line in each 
panel show CMC results with the apocenter criterion and energy criterion, respectively.  
The dotted blue line in the first four panels show the NBODY4 results for the same initial 
clusters for comparison.  The NBODY4 data is taken from the simulations described in 
\citet{2003MNRAS.340..227B}.  Similar simulations using NBODY4 do not exist for 
a higher $N$.  In each case the energy criterion loses mass at a faster 
rate than the apocenter criterion.  When available, the results using the energy 
criterion agree better with the direct $N$-body results.  }
\label{plot:tide}
\end{center}
\end{figure} 

Globular clusters are not isolated systems, but are in fact subject to the tidal
field of their host galaxy.  The assumption of spherical symmetry inherent in MC codes like
CMC does not allow for a direct calculation of stellar loss at the tear-drop shaped tidal
boundary.  Instead, MC codes employ an effective tidal mass loss criterion that attempts to
match the tidal mass loss found in direct $N$-body simulations.  Since stars are lost from
the tidal boundary on a dynamical timescale and MC codes operate on the (much longer) relaxation
timescale, the appropriate effective criterion is not obvious.  There have been two main
suggestions in the literature for the appropriate tidal truncation criterion.  Perhaps
the most natural is to immediately strip any star whose apocenter ($r_a$) lies outside the Roche lobe radius
of the cluster (which we call the tidal radius, $r_t$):
\begin{equation}
r_a > r_t \, .
\label{eq:apo}
\end{equation}
This ``apocenter criterion'' has been used exclusively in CMC previously (Papers I--IV).
Earlier, in the absence of large-$N$ direct $N$-body simulations, comparisons were
made with 2-D Fokker-Plank models and the apocenter criterion showed excellent
agreement (for details see Papers I and II).

Another simple criterion is the ``energy criterion,'' in which any star with a specific orbital energy
above some critical energy is immediately stripped:
\begin{equation}
E_{\rm orb} > 1.5 \phi_t \, ,
\end{equation}
where $\phi_t$ is the cluster potential at the tidal radius \citep{1987degc.book.....S}.
However, a stellar orbit that instantaneously satisfies the above criterion may still remain
bound if it is scattered back to a lower energy orbit before it can escape.
To account for this effect a less
obvious but empirically validated correction factor to the energy criterion above is
suggested by \citet{2008MNRAS.388..429G}:
\begin{equation}
 E_{\rm orb} > \alpha \phi_t \, ,
\label{eq:energy}
\end{equation}
where $\alpha$ is an $N$-dependent parameter given by
\begin{equation}
\alpha = 1.5 - 3\left[\frac{ln(\gamma N)}{N}\right]^{1/4}\, .
\end{equation}

We have re-examined and tested these two criteria (Equations\ \ref{eq:apo} and \ref{eq:energy})
for tidal stellar loss to determine which one
agrees better with the latest results from direct $N$-body simulations.
\citet{2003MNRAS.340..227B} study in detail the tidal dissolution timescales of a cluster in
a tidal field varying the initial number of stars and the initial mass of the cluster.
We have repeated a large subset of this extensive
study of tidal disruption using CMC with both tidal truncation criteria, and compared the
results.

We followed the exact same prescription for setting up the initial conditions
of the clusters as described in \citet{2003MNRAS.340..227B}.
Each cluster in this set
of runs is assumed to have a circular orbit around the Galactic center with radius $8.5\,\rm{kpc}$.
The Galactocentric circular
velocity $V_G$ is assumed to be $220\,\rm{km/s}$.  The initial position and velocity of
each star of the cluster is chosen from a King model distribution function with central concentration
parameter $W_0 = 7$.  The initial masses of the stars are chosen according to the
Kroupa 2001 mass function in the range $0.1-15\,\rm{M_\odot}$.  We vary the initial
number of stars in the clusters between $16000$ and $10^6$.  There are no primordial
binaries in these simulations.

Figure\ \ref{plot:tide} shows the evolution of the bound mass in a cluster
in the standard Galactic tidal field as described above.
We find that the energy criterion results agree better with the direct $N$-body results,
when available from
\citet{2003MNRAS.340..227B}. The agreement is much poorer when using the apocenter criterion.
For example, for the direct $N$-body run with initial $N = 64K$, the time when the cluster
loses $80\%$ of its initial mass is $\approx 11\,\rm{Gyr}$.  Using CMC with the energy
criterion this value is $\approx 10\,\rm{Gyr}$, whereas, with the apocenter criterion the
same cluster does not lose $80\%$ of its initial mass within $20\,\rm{Gyr}$, the integration
stopping time.

Since the energy criterion gives a significantly better
agreement with existing direct $N$-body results (Figure\ \ref{plot:tide}; see also more detailed
comparisons in \S\ref{comparison}), we adopt the energy criterion here,
in contrast to what was used in our earlier works (Papers II -- IV), unless otherwise mentioned.

One should note, however, that the cluster mass range where direct
$N$-body results are available is not
representative of the actual GGC population.  Since no
direct $N$-body results exist for more massive clusters, it is not possible
at present to determine which approximation is more accurate for larger $N$.

The two  criteria above are treated as initial options in CMC and either one can be selected
at the beginning of a simulation.  At each timestep, the amount of mass lost is calculated using the
chosen criterion in an iterative way to obtain the bound mass (see Paper II for details).

\section{Comparison with direct $N$-body Results}
\label{comparison}
In this section we validate our treatment of stellar evolution by comparing with results 
from previously published studies using the popular direct $N$-body code NBODY4 \citep{2003gnbs.book.....A}.  Since the direct $N$-body simulations employ the least 
degree of assumptions, we tend to trust them more for validation purposes.    

One of the biggest simulations treating all relevant physical effects including primordial 
binaries and stellar evolution was performed by 
\citet{2007ApJ...665..707H, 2007MNRAS.379...93H}.  In particular they studied 
the evolution of the core properties, binary number fractions in the core as well as in the full 
cluster, and the evolution of the bound number of stars.  Since both these works present 
data from a common set of simulations we henceforth collectively call them Hurley07.    

\subsection{Initial Conditions}
\label{hurley_ic}
We choose from Hurley07 the simulations with a number $N_i = 10^5$ initial objects 
(the largest initial $N_i$ in their set of simulations), with primordial 
binary fractions $f_{b}=5\%$ and $f_{b}=10\%$
(their K100-5 and K100-10 models, respectively).  
Throughout this work we count binary center of mass as one object.  Thus a cluster with 
$N_i=10^5$ and $f_{b,i}=5\%$ initially has $95000$ single stars and $5000$ binaries.  
We simulate clusters with exactly the same initial conditions using CMC.  The initial stellar 
positions and velocities are chosen from a virialized Plummer sphere.  The stellar masses 
are chosen from the IMF presented in \citet[][henceforth K91]{1991MNRAS.251..293K} 
in the range $0.1-50\,\rm{M_\odot}$.  Metallicity is fixed at $z=0.001$.  Each cluster has 
an initial virial radius $r_v=8.5\,\rm{pc}$ which corresponds to a roche-filling cluster with 
tidal radius 
$r_t \sim 50\,\rm{pc}$, consistent with the standard Galactic tidal field with a Galactic 
rotation speed $220\,\rm{km/s}$ at a Galactocentric distance $8.5\,\rm{kpc}$.  The companion 
mass in each binary is chosen from a uniform distribution in mass ratio in the range 
$0.1-m_p\,\rm{M_\odot}$, where, $m_p$ is the primary mass.  The binary period is drawn 
from a distribution flat in logarithmic intervals in the semi-major axes \citep{1989ApJ...347..998E} and the eccentricities are thermal 
following Hurley07.  We call these simulations {\tt hcn1e5b5} and {\tt hcn1e5b10}, 
respectively (Table\ \ref{tab:hc}).                
\begin{deluxetable}{cccccccc}
\tabletypesize{\footnotesize}
\tablecolumns{8}
\tablewidth{0pt}
\tablecaption{Initial conditions for comparison runs with Hurley07 \label{tab:hc}}
\tablehead{
\colhead{Name} & 
\colhead{N} & 
\colhead{M ($10^4\rm{M_\odot}$)} & 
\colhead{Profile} & 
\colhead{IMF} & 
\colhead{$r_{t}$ (pc)} & 
\colhead{$r_{v}$ (pc)} & 
\colhead{$f_{b}$}}
\startdata
hcn1e5b5 & $10^5$ & $5$ & Plummer & K91\tablenotemark{a} $[0.1, 50]\,\rm{M_\odot}$ & $51$ & $8.5$ & $0.05$ \\
hcn1e5b10 & $10^5$ & $5$ & Plummer & K91 $[0.1, 50]\,\rm{M_\odot}$ & $51$ & $8.5$ & $0.1$ \\
\enddata
\tablenotetext{a}{Kroupa mass function as described in \citet{1991MNRAS.251..293K}.  }
\end{deluxetable}
%
\subsection{Comparison of Results}
\label{hurley_results}
\begin{figure}
\begin{center}
\plotone{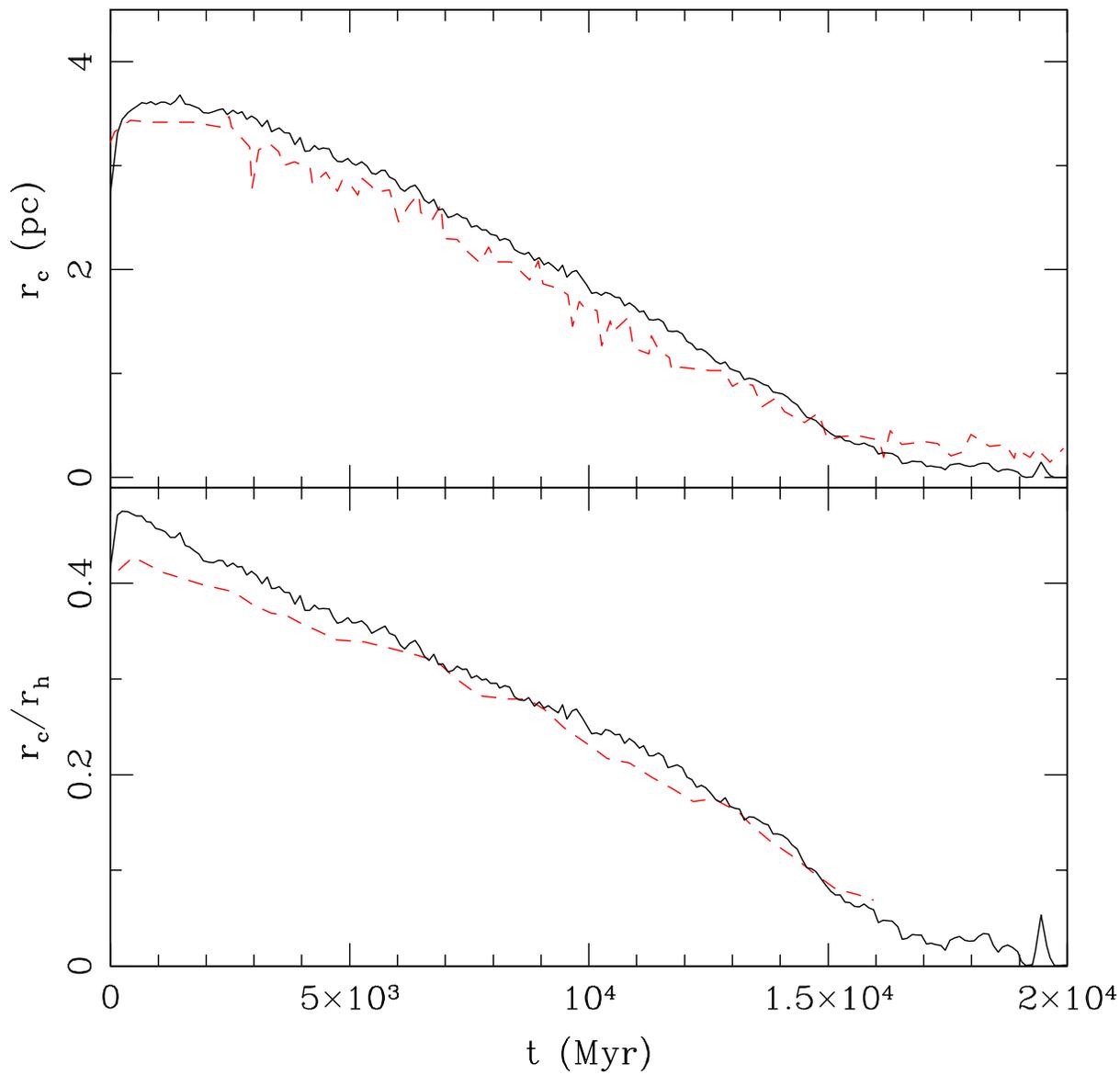}
\caption{Evolution of $r_c$ (top) and $r_c/r_h$ (bottom).  The solid black lines show 
results from simulation using CMC and the red dashed lines show results from Hurley07.  
All data from Hurley07 used for comparison in this work are extracted using ADS's 
Dexter data extraction applet \citep{2001ASPC..238..321D}.  
}
\label{plot:b5_rc}
\end{center}
\end{figure} 
For the global evolution of a dense cluster the evolution of the core is 
extremely important since throughout the evolution the global properties of 
the cluster are determined by the balance of energy in and out of the core.  
The core radius ($r_c$) is also one of the most easily observable structural 
properties of a cluster.  
Moreover, the properties and the evolution of $r_c$ is one of the easiest theoretical 
way to characterize the distinct phases of a cluster's evolution.  
Thus a basic test for the validity of a cluster simulation is to compare 
the evolution of the core radius ($r_c$) and the ratio of the core to half mass radius 
($r_c/r_h$).  

Note that for all our simulated clusters in this work $r_c$ is the density-weighted 
core radius \citep{1985ApJ...298...80C} commonly used in $N$-body simulations, 
unless otherwise specified.  This is not a directly observable quantity and 
can differ from the observed $r_c$ by a factor of a few \citep{2007MNRAS.379...93H}.     

Figure\ \ref{plot:b5_rc} shows the evolution of $r_c$ for 
run {\tt hcn1e5b5} (Table\ \ref{tab:hc}) and K100-5 in Hurley07.  
The scale-free quantity $r_c/r_h$ is also 
plotted for each run.  The core radius expands due to stellar evolution mass 
loss during the first 
$\sim 2\times10^2\,\rm{Myr}$.  The core then contracts at a steady rate till a little after 
$\sim 1.5\times10^4\,\rm{Myr}$.  The core radius then attains a relatively steadier value 
as the cluster reaches the binary-burning phase \citep[e.g.][]{2007ApJ...658.1047F}.  
All these qualitatively different phases of the evolution of a cluster is reproduced 
using CMC with excellent agreement.

\begin{figure}
\begin{center}
\plotone{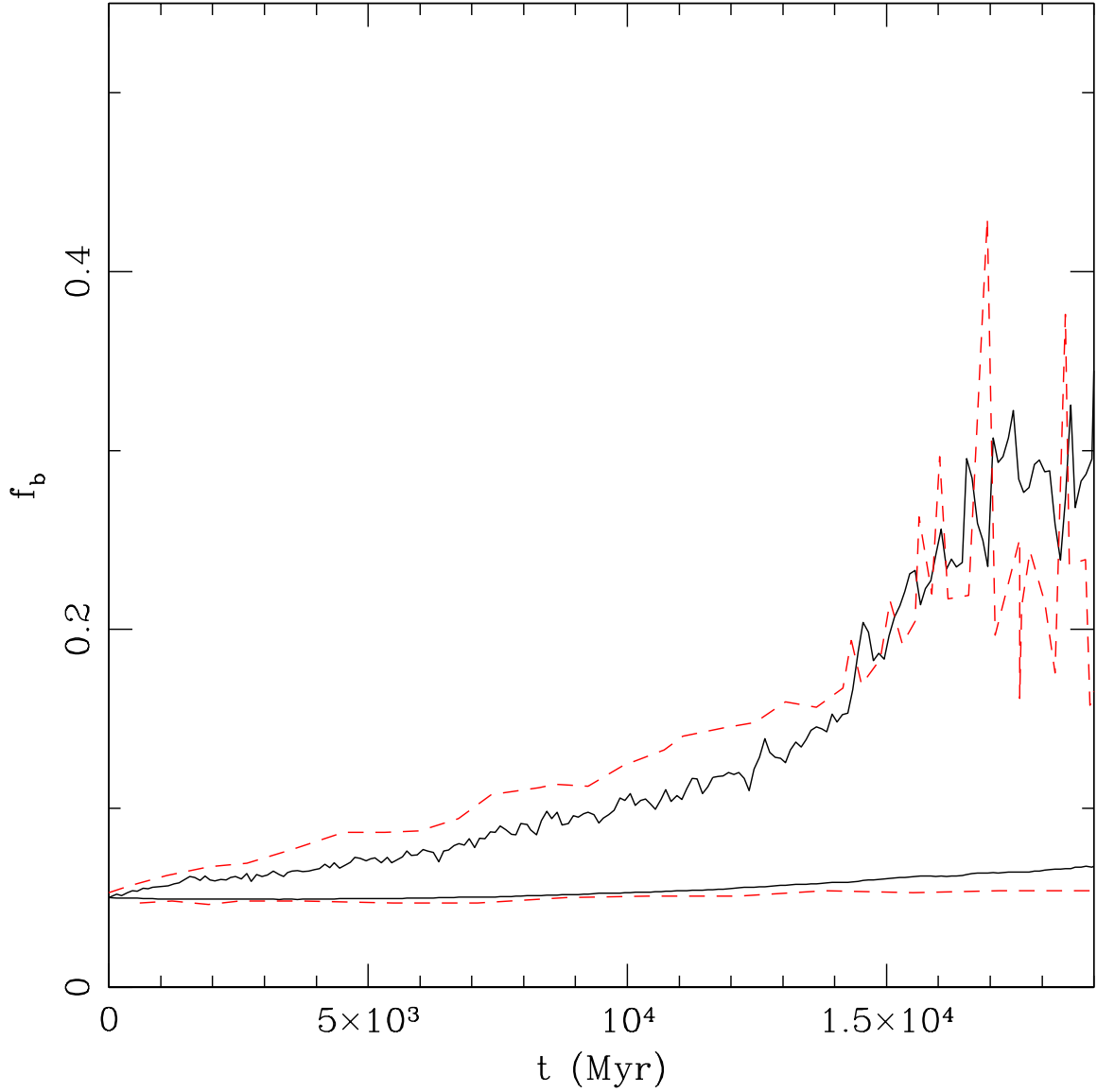}
\caption{Comparison of evolution of the core and the overall binary fraction.   
Dashed red lines show results from \citet[][their K100-5 run]{2007ApJ...665..707H}.  
Solid black lines show results from CMC run using the exact same initial cluster 
parameters.  In both sets the top line show the binary fraction within the core, and the 
bottom line show that for the whole cluster.  }  
\label{plot:b5_fb}
\end{center}
\end{figure} 
\begin{figure}
\begin{center}
\plotone{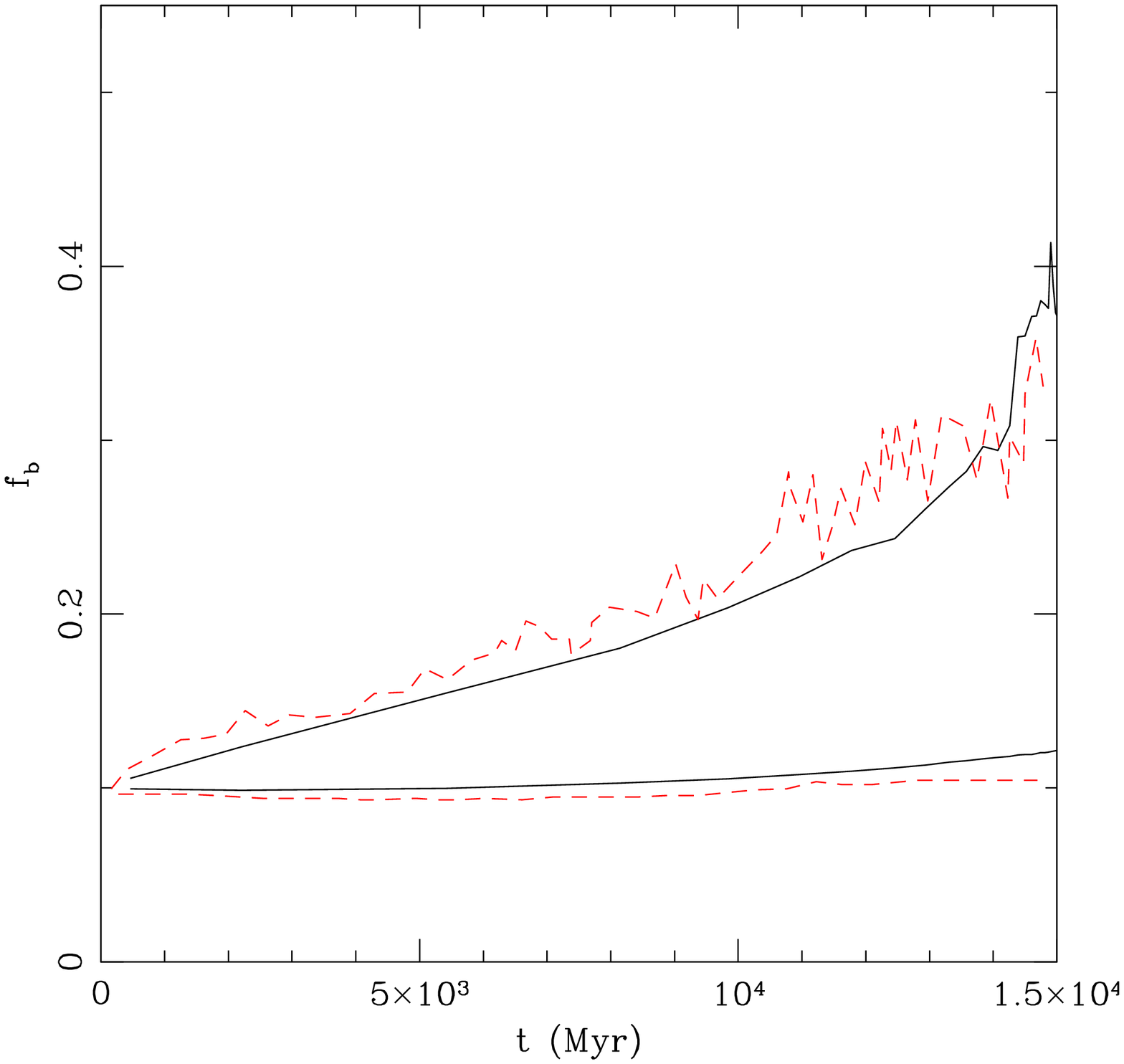}
\caption{Same as Figure\,\ref{plot:b5_fb}, using results from simulation {\tt hcn1e5b10}.  The 
CMC results are compared with the K100-10 simulation of Hurley07.   }
\label{plot:b10_fb}
\end{center}
\end{figure} 
One of the key results of Hurley07 is that the overall binary fraction ($f_b$) remains close 
to the primordial value throughout the evolution of the cluster 
\citep[also see][]{2009arXiv0907.4196F}.  
This result has immense observational significance.  In practice only the present day 
properties of a cluster are observed.  This result from Hurley07 indicates that if a 
present day binary fraction of the cluster close to $r_h$ can be observed the 
primordial hard $f_b$ should have been close to this observed value.  
Figure\ \ref{plot:b5_fb} shows the evolution of the core ($f_{b,c}$) and the overall $f_b$ 
from CMC simulation {\tt hcn1e5b5} and direct $N$-body simulation presented in Hurley07.  
Binaries preferentially sink to the center due to mass segregation 
and the single stars typically get tidally disrupted from the tidal boundary.  These two effects 
compete with each other--- the first reduces and the second increases $f_b$ outside the core.  
For the simulated cluster these two effects more or less 
balance each other.  For the simulated cluster 
we reproduce the results presented in Hurley07 and verify that the overall 
$f_b$ remains close to the primordial value whereas $f_{b,c}$ increases over time.  
Similar results are found for the simulation {\tt hcn1e5b10} (Figure\ \ref{plot:b10_fb}).  

\begin{figure}
\begin{center}
\plotone{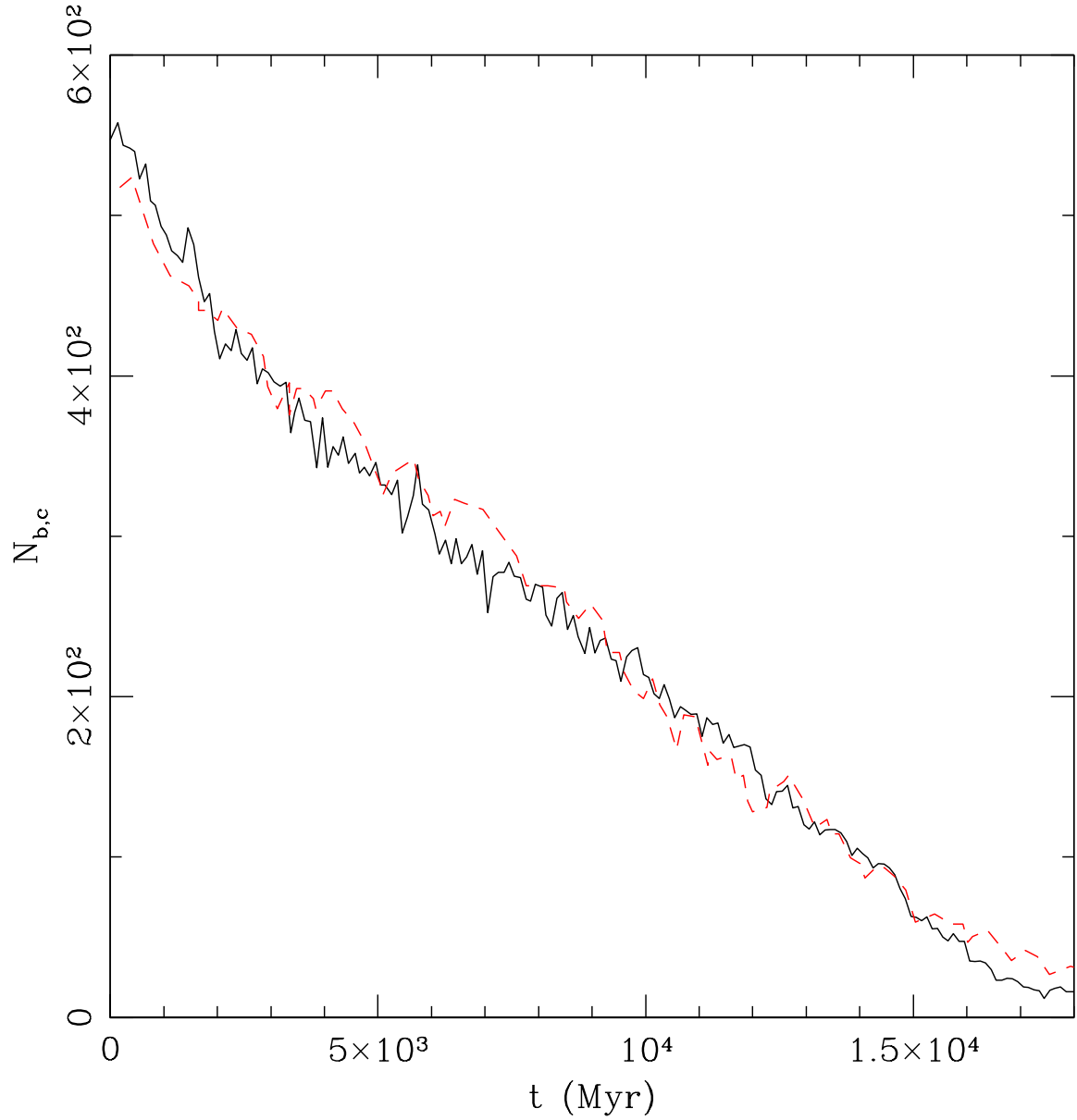}
\caption{Evolution of number of binaries within $r_c$ ($N_{b,c}$).  
The solid black line shows 
the results from CMC run {\tt hcn1e5b5} and the red dashed line shows results from 
direct $N$-body run K100-5 from Hurley07.  }
\label{plot:hc_Nbc}
\end{center}
\end{figure} 
\begin{figure}
\begin{center}
\plotone{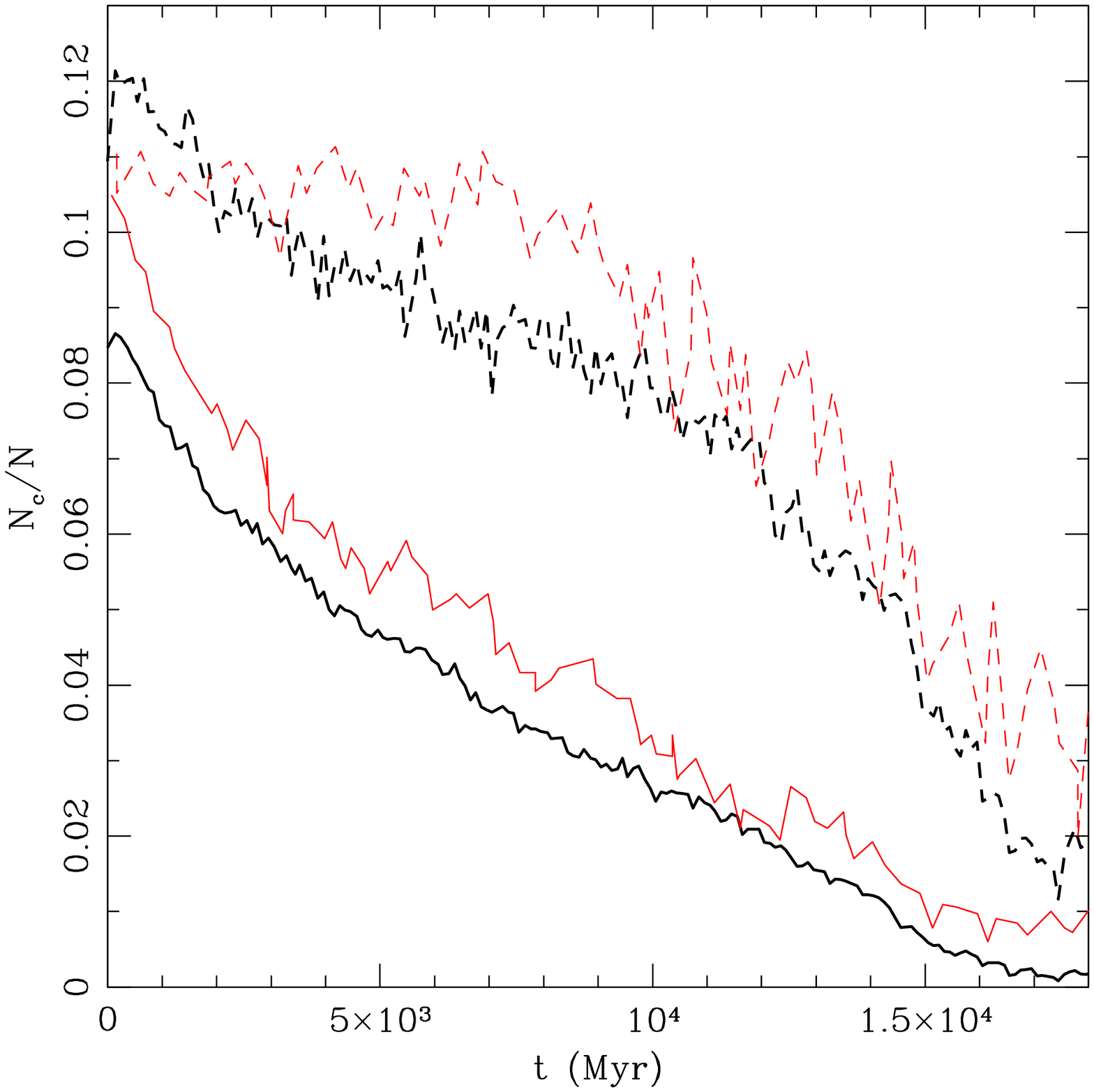}
\caption{Evolution of the fraction (by numbers) of binary and single stars within the core.  Thick black lines show the results 
from CMC.  Thin red lines show the same from Hurley07.  Solid and dashed lines show the number of singles and binaries within $r_c$, respectively.  All 
numbers are normalized with the total number of that species at that time in the cluster 
(e.g., $N_{b,c}/N_b$ for the binaries).  
Both results clearly show the effects of mass segregation since throughout the 
evolution a higher fraction of binaries reside in the core.  }
\label{plot:hc_nc}
\end{center}
\end{figure} 
We now focus on the evolution of the number of binaries in the core ($N_{b,c}$).  The 
evolution of the total number of core binaries is interesting for various reasons.  The formation 
rates of interesting stellar objects such as X-ray binaries and blue straggler stars 
and their properties are directly dependent on $N_{b,c}$, motivating many detailed 
studies focusing on the evolution of $N_{b,c}$ 
\citep[e.g.][]{2002MNRAS.329..897H,2005MNRAS.358..572I,2006MNRAS.372.1043I,2008MNRAS.386..553I,2009arXiv0907.4196F}.   
On one hand the core binary number ($N_{b,c}$) increases due to mass segregation.  
On the other hand strong interactions involving binary-single (BS) and binary-binary (BB) 
encounters can lead to direct physical collisions or destruction of soft binaries and reduce 
$N_{b,c}$.  In addition, binary stellar evolution can destroy binaries via evolution-driven 
mergers and disruptions.  Since the evolution of $N_{b,c}$ is dependent on these competing 
effects it is not simple to predict its evolution a-priori.  Figure\ \ref{plot:hc_Nbc} shows the 
evolution of the number of core binaries ($N_{b,c}$) for CMC run 
{\tt hcn1e5b5} and direct $N$-body 
run K100-5 from Hurley07 for comparison.  
The evolution of $N_{b,c}$ is reproduced exactly within the numerical fluctuations 
of the simulations.  
Over time the number of core 
binaries ($N_{b,c}$) decreases.  

It is also interesting to study the number fraction of binaries and single stars within the 
core compared to the global population.     
Although $N_{b,c}$ decreases over time, due to mass segregation effects the 
number of single stars within $r_c$ decreases more.
Figure\ \ref{plot:hc_nc} shows the evolution of the number fractions of single stars 
($n_{s,c}$) and binaries ($n_{b,c}$) within $r_c$ for the same simulations as above.  
During the first $\sim 10^4\,\rm{Myr}$ $n_{b,c}$ remains more or less constant whereas, 
$n_{s,c}$ decreases by $\sim 0.5$ of the initial $n_{s,c}$ due to mass segregation effects.  
Followed by this phase BS/BB interactions as well as stellar evolution destroys core 
binaries decreasing $n_{b,c}$.  However, throughout the evolution $n_{b,c}>n_{s,c}$.  
The combined effects of the above leads to the overall increase in $f_{b,c}$ over time 
as seen in Figures\ \ref{plot:b5_fb} and \ref{plot:b10_fb}.  
Note that although qualitatively CMC results and the direct $N$-body results agree, the 
agreement is not as excellent as the previous comparisons.  For example, for the evolution 
of $n_{b,c}$ there can be upto $\approx 20\%$ difference in the absolute value depending 
on the age of the simulated cluster.  
The reason behind this larger difference compared to the excellent agreement for the 
evolution of $N_{b,c}$ (Figure\ \ref{plot:hc_Nbc}) originates from the approximations 
adopted in the tidal treatment in MC methods.  
\begin{figure}
\begin{center}
\plotone{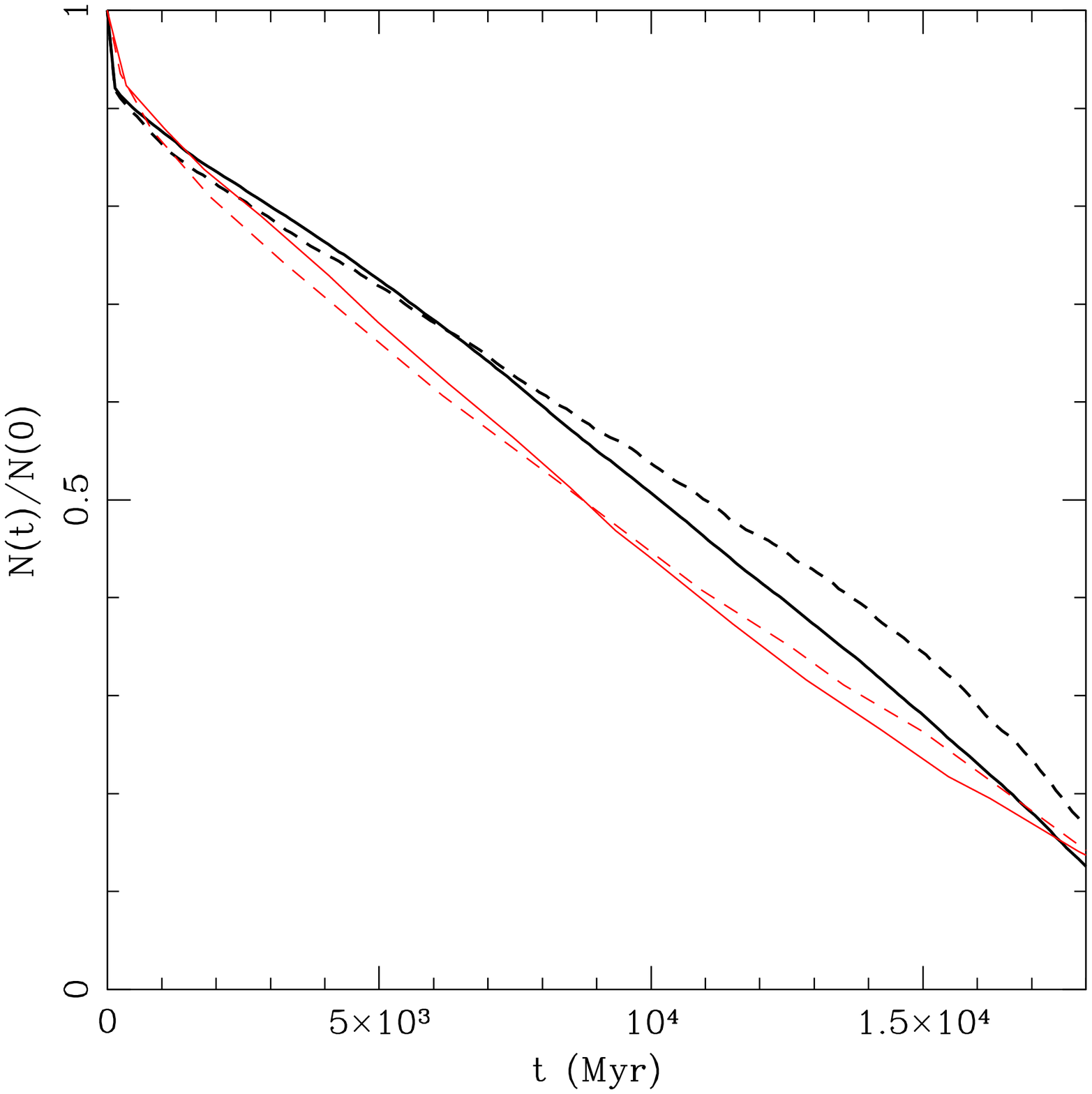}
\caption{Comparison of the evolution of the number of single and binary stars that 
remain bound to the cluster.  
Thinner lines show results obtained from Hurley07.  Thicker lines show results from 
CMC run using the same initial cluster parameters.  
Solid lines for both cases show the number of single stars bound at any given time.  Dashed 
lines show the same for binaries.  All numbers are normalized to the initial number of the same 
species (single or binary).  }
\label{plot:hc_nbound}
\end{center}
\end{figure} 
The criterion based tidal removal of 
stars adopted in CMC loses stars from the tidal boundary at a relatively lower rate 
(Figure\ \ref{plot:hc_nbound}).  Hence, at a given time the total number of bound single 
and binary stars in CMC are higher than those in Hurley07 making 
both $n_{b,c}$ and $n_{s,c}$ calculated using CMC systematically lower than 
the same calculated in Hurley07.    

\begin{figure}
\begin{center}
\plotone{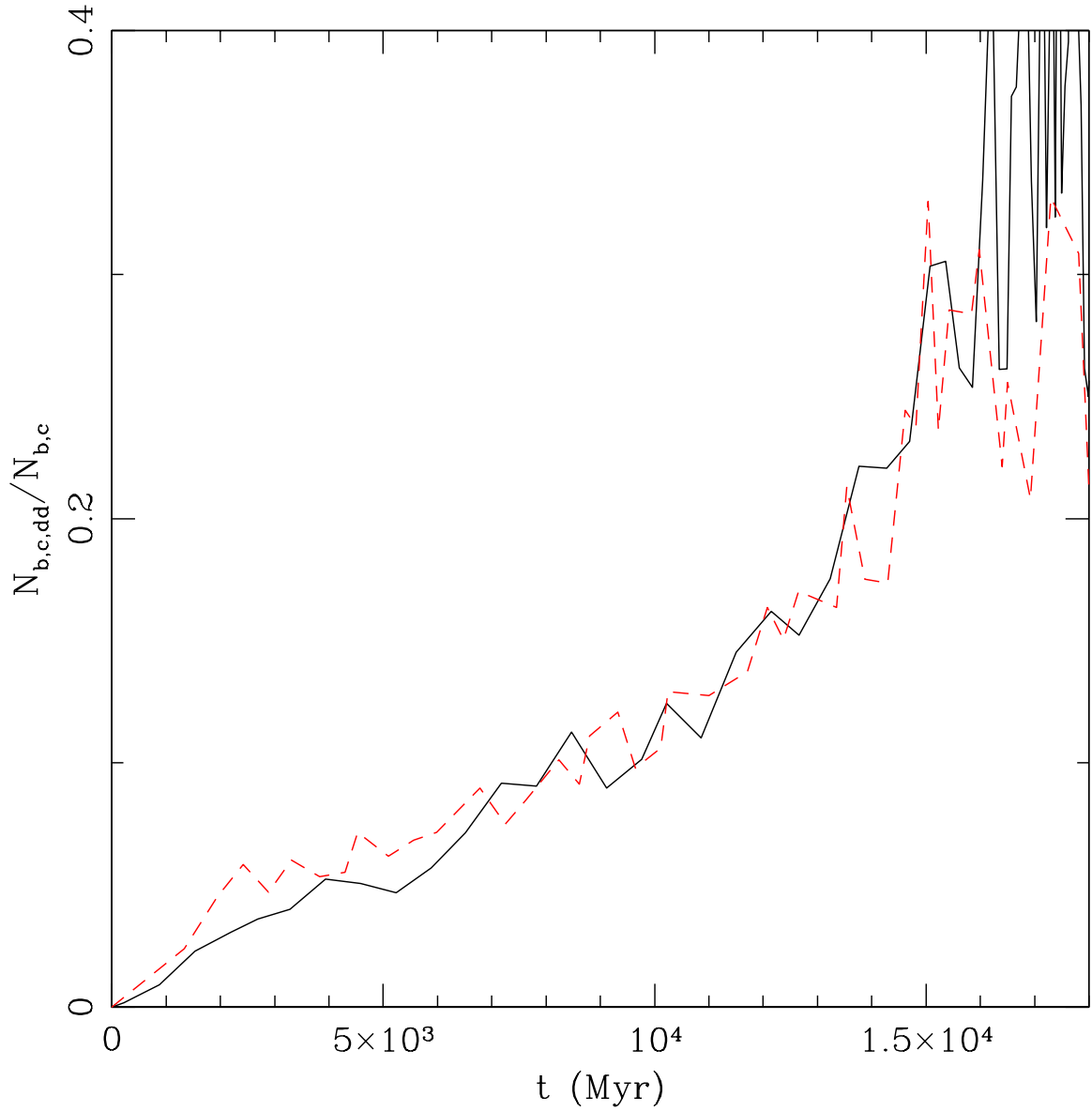}
\caption{Evolution of the fraction of double-degenerate binaries within $r_c$ 
with respect to all core binaries.  Black solid and red dashed lines show results 
from CMC simulation and direct $N$-body simulation in Hurley07, respectively.  }
\label{plot:hc_nbc_dd}
\end{center}
\end{figure} 
Another interesting result presented in Hurley07 is the evolution of the 
fraction of binaries in the core 
where both components are compact objects.  We call them double-degenerate binaries, 
following Hurley07.  
In Figure\ \ref{plot:hc_nbc_dd} we show the evolution of the fraction of 
double-degenerate core binaries 
for CMC run {\tt hcn1e5b5} and direct $N$-body run 
K100-5.  The fraction of double-degerate binaries in the core depends on all physical 
processes in the cluster in a complicated way.  
Two-body relaxation drives 
mass segregation in the cluster determining the densities at different radial regions of the 
cluster as well as radius dependent velocity dispersion.  
This in turn directly affects the local BS/BB scattering cross-section at a given 
time consequently determining the survivability of a given binary at some radial position 
in the cluster and also the properties of the stellar members in a binary and the binary 
orbit.    
Changing the binary stellar properties and their orbital properties in turn changes 
the evolutionary pathways taken by the binary members and consequently compact 
object formation.  
Stellar evolution and dynamical effects thus in tandem affect the fraction 
of double-degerate binaries in the core.  
The excellent agreement between CMC results with Hurley07 convinces us that 
not only the dynamical effects, but also the stellar evolution, and the rate of compact 
object formation, are modeled at least as accurately as in the 
direct $N$-body code NBODY4.  

\section{Comparison With Simulations Without Stellar Evolution}
\label{without}
We now examine the effects of stellar evolution on the evolution of 
the global observable properties of a GC, by performing comparisons to 
simulations without stellar evolution.  The initial conditions 
for these simulations are summarized in Table\ \ref{tab:without}.    

\begin{deluxetable}{ccccc}
\tabletypesize{\footnotesize}
\tablecolumns{8}
\tablewidth{0pt}
\tablecaption{Initial conditions for comparison runs including and leaving out stellar evolution \label{tab:without}}
\tablehead{
\colhead{Name} & 
\colhead{N} & 
\colhead{Profile} & 
\colhead{IMF} & 
\colhead{$f_{b}$}}
\startdata
kw4b03 & $10^5$ & King & K01 $[0.1, 1.2]\,\rm{M_\odot}$ & $0.03$ \\
kw4b1 & $10^5$ & King & K01$[0.1, 1.2]\,\rm{M_\odot}$ & $0.1$ \\
kw4b3 & $10^5$ & King & K01 $[0.1, 1.2]\,\rm{M_\odot}$ & $0.3$ \\
kw7b0 & $5\times10^5$ & King & K01 $[0.1, 18.5]\,\rm{M_\odot}$ & $0$ \\
kw7b1 & $5\times10^5$ & King & K01 $[0.1, 18.5]\,\rm{M_\odot}$ & $0.1$ \\
\enddata
\end{deluxetable}

Our most recent earlier paper, Paper IV, showed results from simulations without stellar evolution, but all other physical 
processes were included.  In the absence of an implementation of full single and 
binary stellar evolution Paper IV restricted itself to simulations with a narrow range 
of masses in the IMF.  We first compare the results with stellar evolution with a small subset 
of the previous runs from Paper IV without stellar evolution with the narrow IMF as an example.  
Since this is for the purpose of comparison, we use the apocenter criterion (\S\ref{method}) 
for the tidal treatment to be consistent with Paper IV for these simulations.  

For each of these simulations the initial stellar positions and velocities are chosen from 
a King profile with the concentration parameter $W_0 = 4$.  For each simulation $N_i = 10^5$.  
The stellar IMF is chosen 
from the stellar MF presented in \citet[][henceforth K01]{2001MNRAS.322..231K}.  The 
initial binary fraction is varied between $f_b = 0.03, 0.1$, and $0.3$.  The mass of each 
binary companion is chosen in the range $0.1-m_p$ from a uniform distribution in mass 
ratios.  The binary periods are chosen from a distribution flat in logarithmic $a$ intervals 
within physical limits, 
where, the hardest binary has $a >$ the sum of the stellar radii of the companions and the 
softest binary is at the local hard-soft boundary.  Binary eccentricities are thermal 
\citep[e.g.][]{2003gmbp.book.....H}.   For each 
of these initial conditions one simulation is done including stellar evolution and the other 
leaving it out.    

\begin{figure}
\begin{center}
\plotone{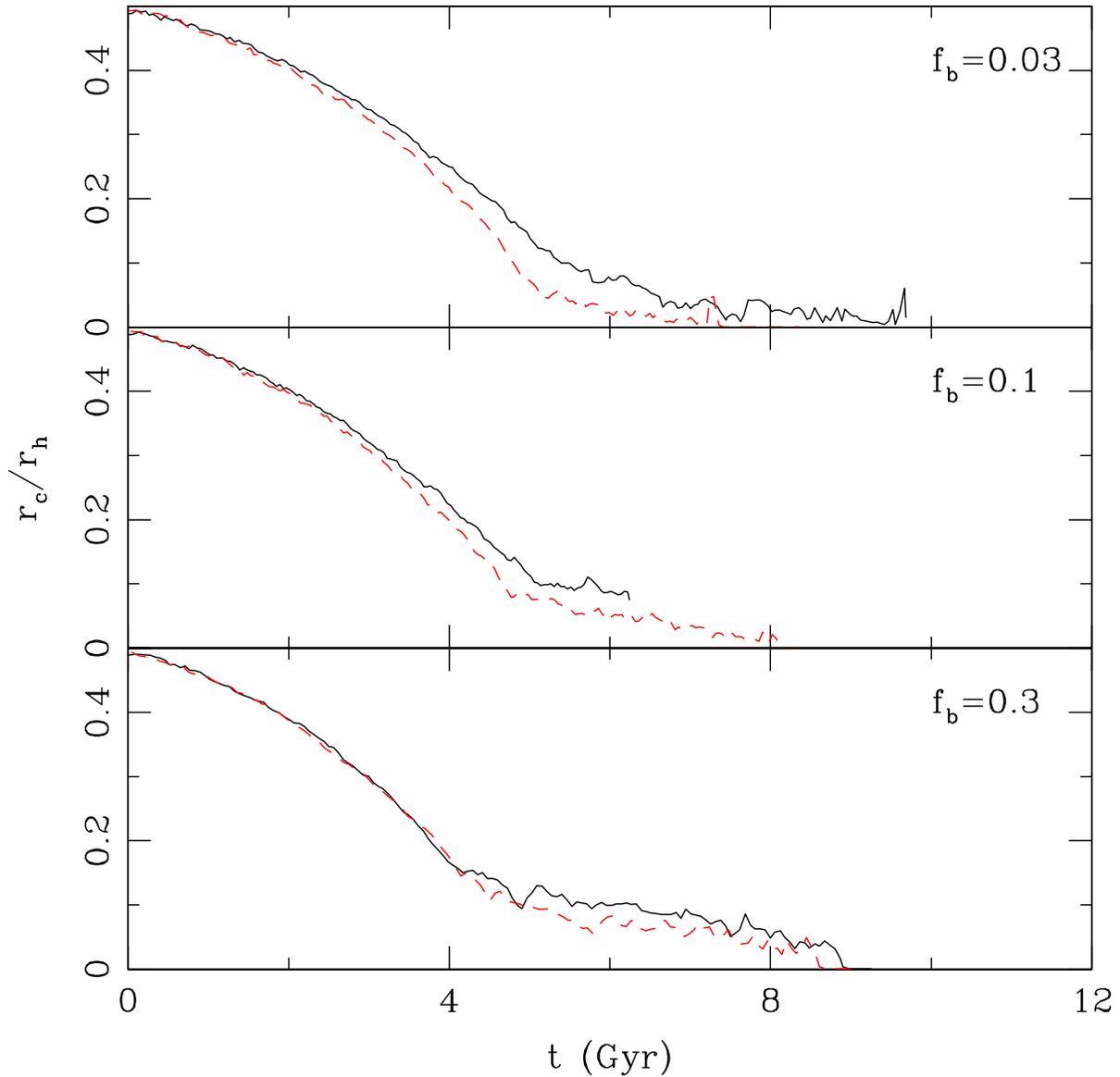}
\caption{Comparison of the evolution of $r_c/r_h$ with (solid black) and without 
(dashed red) stellar evolution, both using CMC.  
Results without stellar evolution were already presented in Paper IV.  
Each run starts with 
$10^5$ objects.  
The velocities and positions 
of the objects are chosen from a King profile with an initial $W_0 = 4$.  The masses are chosen 
from a Salpeter MF in the range $0.2 - 1.2\,\rm{M_\odot}$.  From top to bottom the initial 
binary fractions are $0.03$, $0.1$, and $0.3$.  }
\label{plot:w4n1e5}
\end{center}
\end{figure}

We find that, even for the simulations with a small range of initial stellar masses, where 
the stellar evolution mass loss is not as severe as in a realistic cluster, 
for low $f_b$ stellar evolution can influence the overall cluster evolution to a 
certain extent.    
Figure\ \ref{plot:w4n1e5} shows 
the evolution of $r_c/r_h$.  The results are shown 
for runs {\tt kw4b0.03, 0.1, 0.3} (see Table\ \ref{tab:without}).  From top to bottom 
the primordial binary fractions $f_b$ are $0.03$, $0.1$, $0.3$, respectively.  For 
$f_b = 0.03$ even with the narrow mass range the two curves start diverging when the most 
massive stars (in this case $1.2\,\rm{M_\odot}$) evolve off their MS and lose mass via 
compact object formation after $\approx 3.4\,\rm{Gyr}$.

Binary interactions take place throughout the evolution of the cluster.  As the initial $f_b$ 
increases, energy available from super-elastic scattering of binaries becomes relatively 
more important compared to the energy produced from stellar evolution mass loss.  
Thus for this narrow range of masses as the binary fraction is increased, the difference 
between the results from simulations including stellar evolution and results without 
including stellar evolution reduces.  For example, evolution of the cluster with 
initial $f_b = 30\%$ is very similar with and without stellar evolution taken into account.  
The only difference is that at the quasi-steady binary-burning phase including stellar 
evolution makes $r_c/r_h$ bigger by about $30\%$.     
In each of these clusters the central densities are not very high 
($\sim 10^4\,\rm{M_\odot/pc^3}$) so direct SS collisions are 
not dominant.  When direct SS collisions are more important in a much denser 
cluster, this behavior may change \citep{2008IAUS..246..151C}.

The difference in the evolution of the global properties depending on whether stellar 
evolution was included or not is, of course, a lot more dramatic when a more realistic IMF 
with a wider mass range is used.  Here we use a 
King profile with central concentration parameter $W_0 = 7$.  The IMF 
is according to the K01 stellar MF in the range $0.1-18.5\,\rm{M_\odot}$.  
Two such clusters are simulated one with no primordial binaries and the other with  
$f_b = 0.1$.  
Note the dramatic difference in the evolution of the simulated clusters.  
\begin{figure}
\begin{center}
\plotone{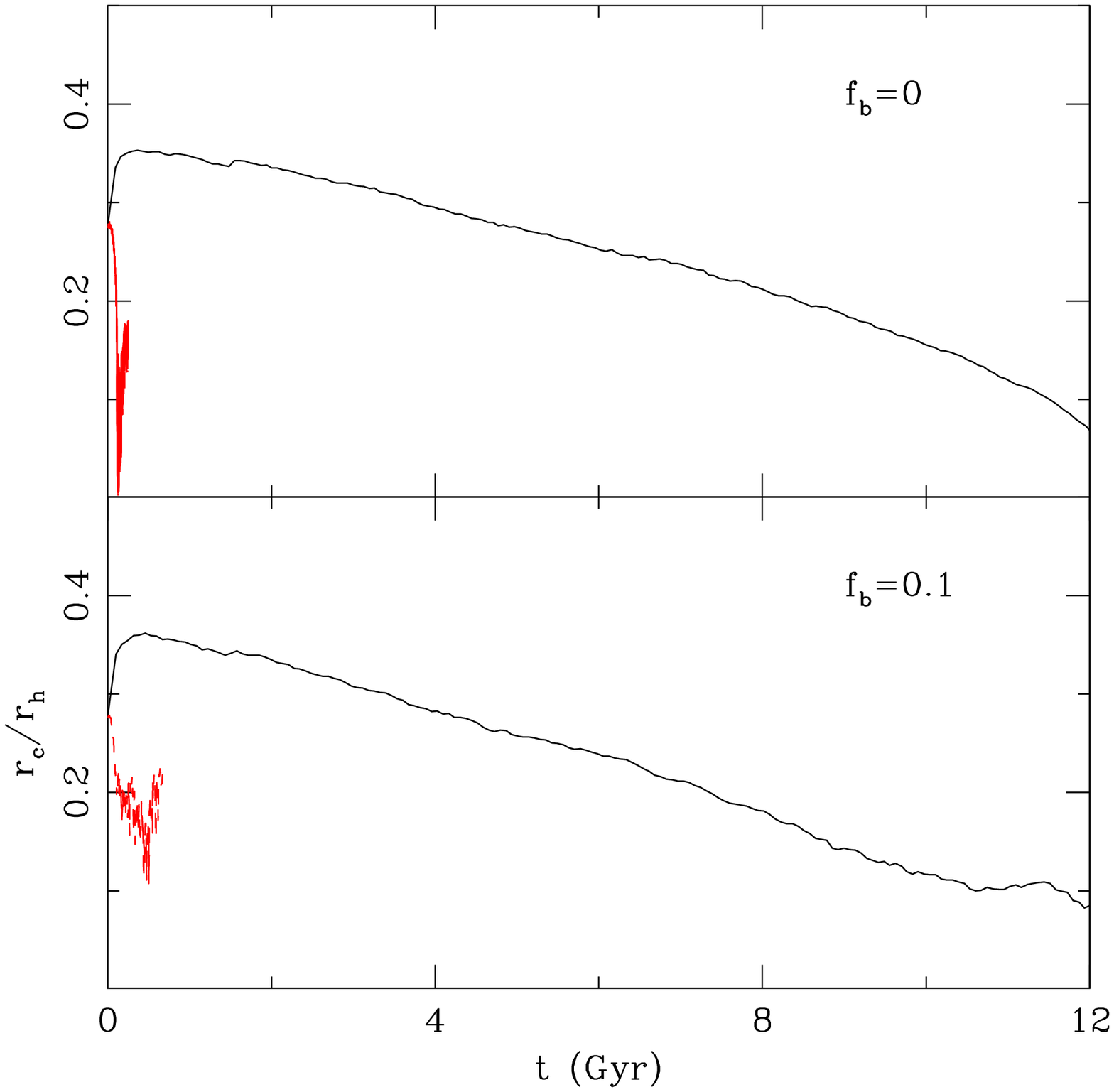}
\caption{Comparison between two sets of simulations, all using CMC, but in one set stellar 
evolution is included and in the other it is not.  
Both panels show simulations with an initial 
King profile cluster with $W_0 = 7$ and a Kroupa 2001 IMF in the range 
$0.1-18.5\,\rm{M_\odot}$.  The top panel shows results with no primordial binaries.  The 
bottom panel shows results with $0.1$ initial $f_b$.  For all simulations the initial number 
of objects is $5\times10^5$.  On both panels dashed red lines are for simulations leaving out 
stellar evolution and solid black lines are for simulations including stellar evolution.  
 A dramatic difference is clearly noticeable caused by stellar evolution mass loss.  }
\label{plot:senose}
\end{center}
\end{figure} 
\begin{figure}
\begin{center}
\plotone{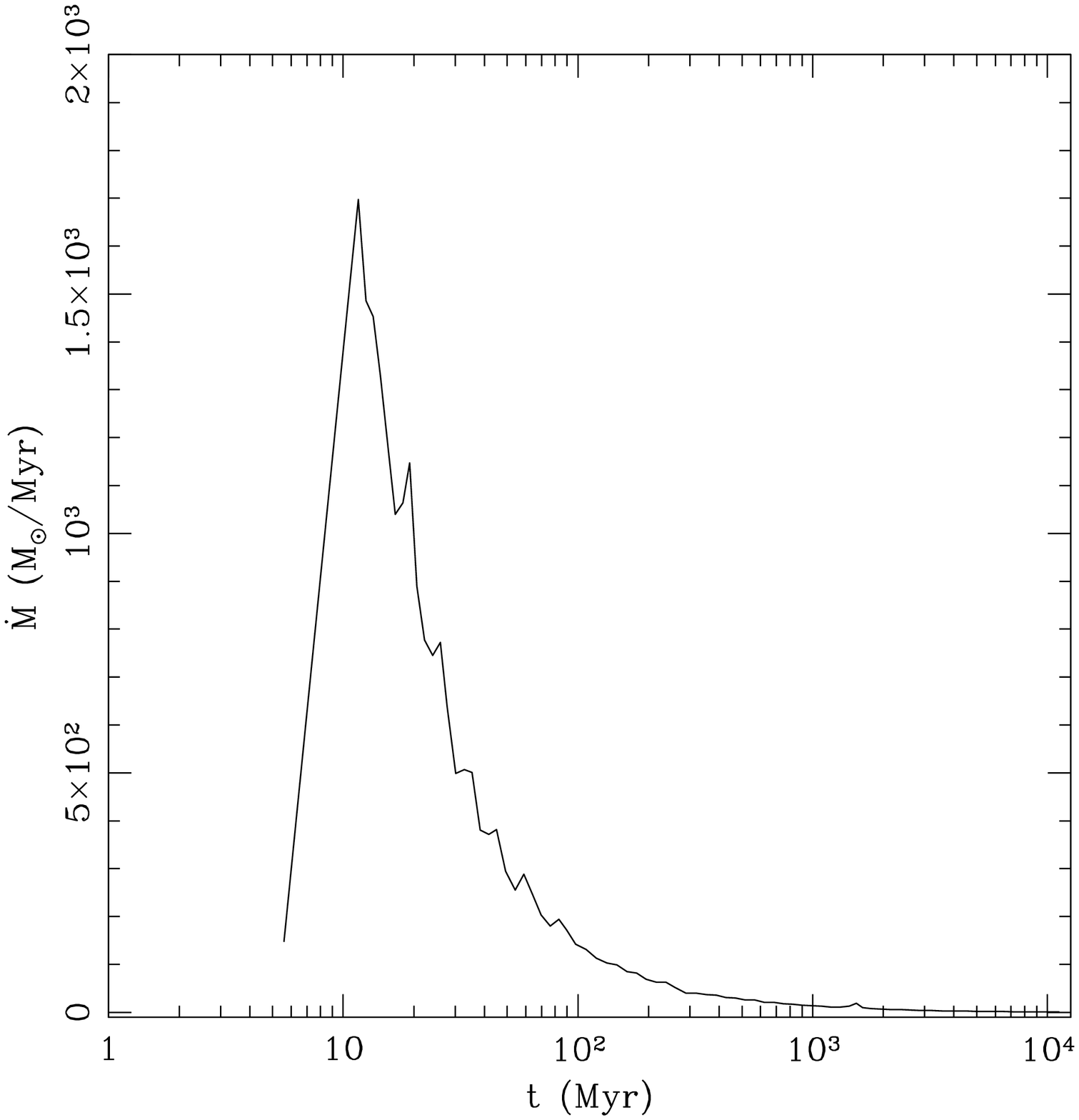}
\caption{Rate of stellar evolution mass loss as a function of time.  Within $10$-$100\,\rm{Myr}$ 
most of the mass is lost due to stellar evolution while high mass stars 
$M_\star \gtrsim 5\,\rm{M_\odot}$ evolve off their MS.  Followed by this initial phase 
the mass loss rate from winds and compact object formation is low.  }
\label{plot:semassloss}
\end{center}
\end{figure} 
The evolution 
during the initial $\sim 10^2\,\rm{Myr}$ is dominated by the mass loss via stellar evolution 
of the high mass stars and 
compact object formation (Figure\ \ref{plot:semassloss}).  
This phase is clearly distinguished by the 
initial steep expansion of the cluster (Figure\ \ref{plot:senose}).  
This phase is followed by a slow contraction phase.  
In this phase, two-body relaxation drives the evolution.  The high mass stars have already 
evolved off the MS and the stars remaining in the cluster are evolving at a much slower 
rate.  The transition between the initial stellar evolution driven expansion and the slow 
contraction happens when the 
energy generation rate from stellar evolution mass loss becomes less than the outwards 
energy diffusion rate from the core due to relaxation.  
The cluster then keeps contracting until the central density increases so much that 
BS/BB interaction rates become high enough and the energy injected by the hard 
binaries (via super-elastic scattering) balances the energy diffusion rate from relaxation.  
The cluster then reaches the binary-burning phase.  
All these phases are clearly seen in Figure\ \ref{plot:senose}, bottom 
panel.  

Even with this moderately broad range in mass, the clusters without stellar evolution 
contracts rapidly and are driven towards a quick collapse.  If there are primordial 
binaries, the binary-burning phase starts relatively early 
($\sim 1\,\rm{Gyr}$, Figure\ \ref{plot:senose}).  On the other 
hand, when stellar evolution is included, even without any primordial binaries 
the same cluster may still be in the slow contraction phase at Hubble time.  For the 
cluster with primordial binaries in this case the binary-burning starts only after 
$11\,\rm{Gyr}$ (Figure\ \ref{plot:senose}).      

\section{Results for Realistic Galactic Globular Clusters}
\label{real}
We have validated CMC by extensive comparisons with direct $N$-body results 
(\S\ref{comparison}).  Moreover, we have shown the importance of including stellar evolution 
in cluster modeling including a realistic stellar IMF (\S\ref{without}).  We now 
simulate a large grid of clusters with realistic initial  
conditions for $12\,\rm{Gyr}$ taking all physical processes into account, 
including primordial binaries and single and binary stellar evolution, 
and the full observed stellar mass range spanning three orders of magnitude.  
Our goal here is to simulate clusters with realistic initial conditions motivated from 
observations of young clusters \citep[e.g.,][]{2007A&A...469..925S} and find whether 
at a simulated cluster age of $\approx 12\,\rm{Gyr}$, a typical age for the GGCs, the simulated 
clusters show similar observable properties (e.g., $r_c$, and $r_c/r_h$) as 
the observed population.  
  
The proper initial conditions for the GGCs are uncertain, however.  
Moreover, it is hard to infer uniquely the 
initial conditions from the present day observed cluster properties since the observed 
cluster global properties as well as their galactic orbits can be quite uncertain 
\citep[e.g.,][]{2008MNRAS.389.1858H}.  
Hence, rather than trying to create a detailed model for any particular cluster we compare the 
collective results of all our grid runs with the observed GGC properties as a whole.   
For comparison the GGC properties are extracted from 
the Harris Catalogue for GGCs 
\citep[][and the references therein; also see http://www.physics.mcmaster.ca/Globular.html]{1996AJ....112.1487H}. 
When an observable is not reported in the catalogue for a cluster, we exclude that cluster from 
comparison.  In the following subsections we explain the initial setup of the grid of simulations 
and present our results.  

\subsection{Initial Conditions}
\label{real_ic}
We simulate clusters with a large grid of initial conditions.  All simulated clusters 
has a fixed initial virial radius $r_v=4\,\rm{pc}$ 
(corresponds to an initial $r_h\approx3\,\rm{pc}$).  Observations indicate that 
the effective radius of both young and old clusters are rather insensitive to the cluster 
mass, and metallicity \citep[e.g.][]{2001AJ....122.1888A,2007A&A...469..925S,2009gcgg.book..103S} and has a median value of $\sim 3\,\rm{pc}$.  In addition, observations of 
old massive LMC clusters, old GCs in NGC\ 5128, old clusters in M\ 51, 
as well as the GGCs indicate 
that the effective cluster radii show only a weak positive relation with 
the distance from the galactic center \citep{1962PASP...74..248H,1984ApJ...287..185H,1984ApJ...276..491H,1987ApJ...323L..41M,1991ApJ...375..594V,2007A&A...469..925S,2008AJ....135.1567H}.             

To restrict the huge parameter space to a certain extent we place all our simulated 
clusters in a circular orbit at a Galacto-centric distance of $r_{GC}=8.5\,\rm{kpc}$, where 
the Galactic field is not so strong that the tidal stellar loss dominates the cluster's evolution. 
Choosing a circular orbit for the simulated clusters is a simplification, however, the results 
should still be valid for eccentric orbits with some effective pericenter distance of $8.5\,\rm{kpc}$ 
\citep[e.g.,][]{2003MNRAS.340..227B}.  
The Galactic tidal field and consequently the initial $r_t$ for the clusters are calculated 
using a Galactic rotation speed $v_G = 220\,\rm{km/s}$ following the standard practice.      

For the set of runs we vary $N_i$ between $4 -10\times10^5$, 
the initial $W_0$ for King models 
in the range $4-7.5$, and initial $f_b$ between $0-0.1$.  
For each case we choose the stellar masses of the primaries 
from the MF presented in K01 in the range $0.1-100\,\rm{M_\odot}$.  
The masses of each binary companion 
is chosen from a uniform distribution of mass ratios in the range $0.1-m_p\,\rm{M_\odot}$.  
$a$ for the binaries are chosen from a 
distribution flat in log within physical limits, namely 
physical contact of the components and the local 
hard-soft boundary.  Although initially each binary is hard at its position it may not remain so during the evolution 
of the cluster.  The cluster contracts under two-body relaxation and the velocity dispersion 
increases making initially hard binaries soft.  Moreover, binaries sink to the core due 
to mass segregation where the velocity dispersion is higher than the velocity dispersion 
for the binaries at $t=0$.  We include these soft binaries in our simulations.  We let the cluster 
dynamics disrupt these binaries via BS/BB interactions.  So at any instant of time soft 
binaries are allowed in the cluster as long as 
they have not been disrupted naturally via dynamical encounters yet.
This is closer to 
reality and this strategy is adopted since soft binaries can act as an energy sink and can 
contribute to the overall cluster energetics significantly \citep{2009arXiv0907.4196F}.  

Each 
cluster is evolved for $12\,\rm{Gyr}$ including all physical processes--- two body relaxation, 
stellar evolution, strong encounters like BB, BS, and SS collisions.  For clusters that reach 
a deep-collapse phase, the CMC time steps become minuscule and the code grinds to a 
halt.  We stop our simulations at that point for these clusters.  Note that in reality, the deep 
collapse phase is halted via formation of the so called three-body binaries and the cluster 
enters into the gravo-thermal oscillation phase.  Since in CMC we do not include the possibility 
of creating new binaries via three body encounters, we do not address this phase at this stage.  
However, this is not a serious limitation for this study since all simulated clusters that reach this 
phase within $12\,\rm{Gyr}$ 
had a primordial $f_b = 0$, which is not realistic 
\citep[e.g., see most recently][]{2008AJ....135.2155D} and simulated for academic interests 
only.
None of the simulated clusters enter into the deep-collapse phase before $\approx 9\,\rm{Gyr}$.  
The properties of all these grid simulations are summarized in Table\ \ref{tab:fixedrv}.      

\subsection{Results}
\label{real_results}
Here we present some basic observable properties of the simulated clusters and compare 
them with the same properties of the observed GGCs.  For each 
of these comparison plots the evolution of a cluster property is shown with the 
distribution of the same property in the GGC population including all GGCs where 
observation of the concerned property exists.  Since we restrict the galacto-centric distance 
of our simulated clusters for this study to be $8.5\,\rm{kpc}$ (\S\ref{real_ic}) we also show 
the observed distribution of the GGCs with pericenter distances from the Galactic 
center within 
$7-10\,\rm{kpc}$ to be consistent in the comparisons.  Note that the purpose for this 
comparison is simply to ensure that the simulated cluster properties agree well with the 
observed GGC properties.  We do not intend to create a present day distribution for these 
properties since for that a probability distribution for the initial conditions is required, which 
is poorly constrained and beyond the scope of this study.  
\begin{figure}
\begin{center}
\plotone{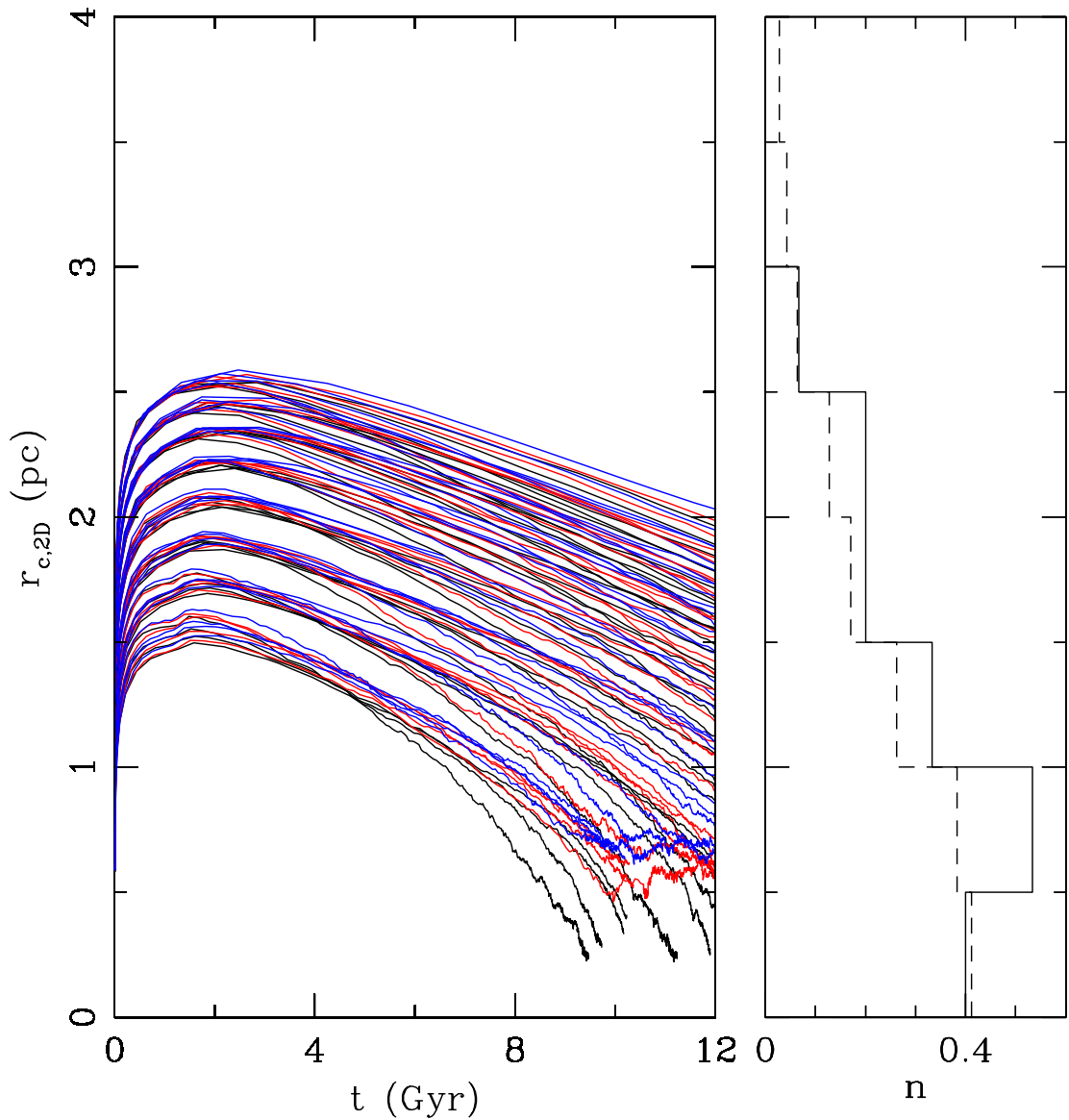}
\caption{Evolution of the 2D projection of $r_c$ for all 
simulated clusters.  The black, red, and blue lines are for clusters with initial 
$f_b = 0$, $0.05$, and $0.1$, respectively.  A few ($6$) clusters with $f_b = 0$ 
go into deep-collapse within a Hubble time.  
We stop integrations for those clusters when this phase is reached.      
The $r_c$ values for the observed GGCs are also shown in histograms.  
The solid histogram is for GGCs with pericenter distances between $7 - 10\,\rm{kpc}$.   
When the orbital eccentricities are unknown, a circular orbit is assumed.  
The dashed histogram is for all GGCs where a measurement for $r_c$ exist.  }
\label{plot:rv_rc_phys}
\end{center}
\end{figure} 
\begin{figure}
\begin{center}
\plotone{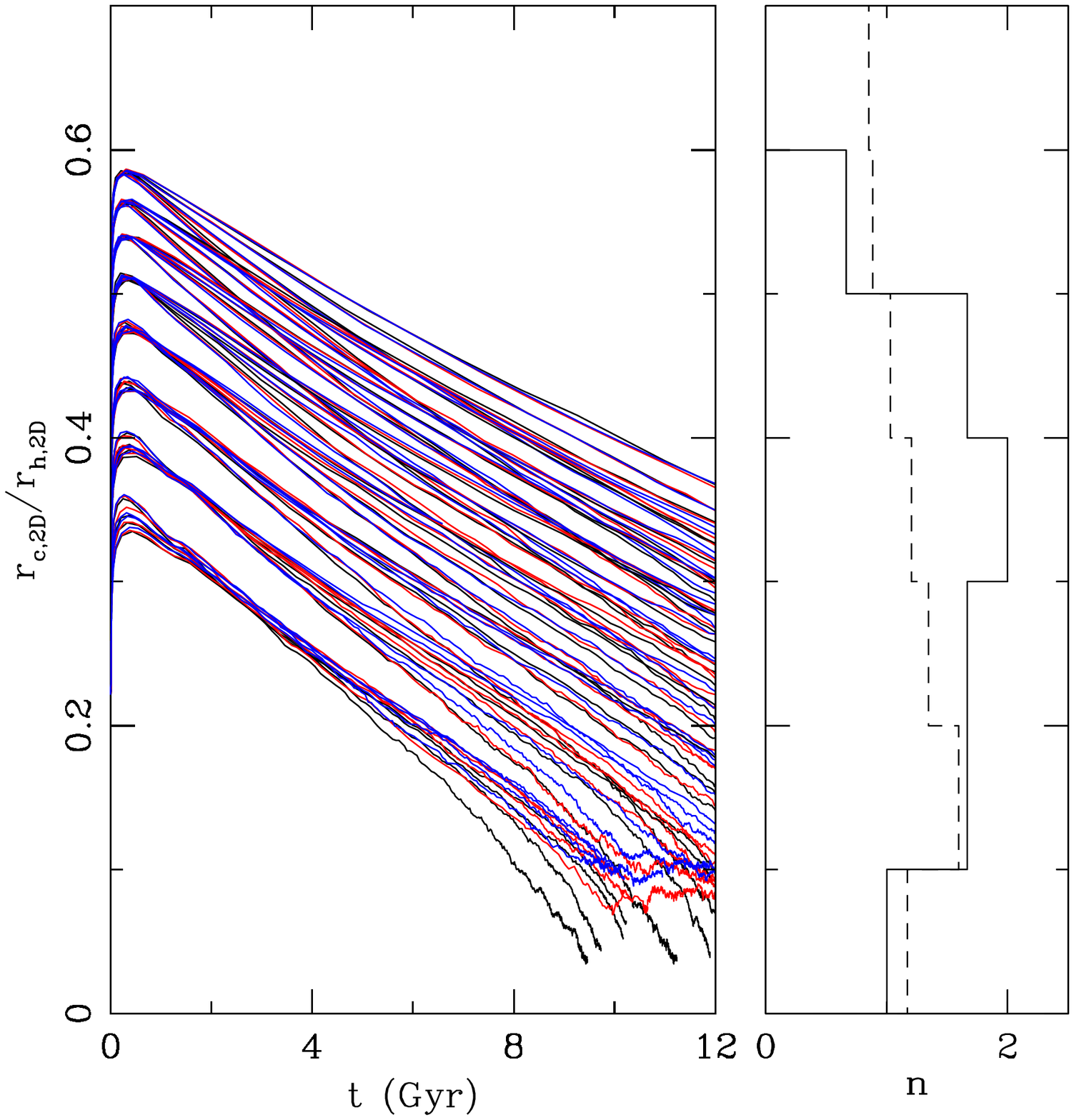}
\caption{Same as Figure\ \ref{plot:rv_rc_phys} for the ratio of the 2D projection 
of the $r_c$ to the 2D projection of the $r_h$ for all simulated clusters.  
The $r_c/r_h$ values for the observed GGCs are also shown in histograms.  
The solid and the dashed histograms are from GGC populations selected as in 
Figure\ \ref{plot:rv_rc_phys}.  
}
\label{plot:rv_rcoverrh}
\end{center}
\end{figure} 

Note that for the observed GGCs only the sky projection of the characteristic radii 
such as $r_c$ and $r_h$ are known.  Hence, in order to be consistent in our comparisons 
with the observed population we show the evolution of the 2D projections of 
$r_c$, and $r_c/r_h$ for all 
simulated clusters (e.g., Figures\ \ref{plot:rv_rc_phys} and \ref{plot:rv_rcoverrh}).  
The sky projections for all simulated clusters are 
done assuming spherical symmetry.    
Both $r_c$ as well as the $r_c/r_h$ values of the simulated clusters agree well with  
the observed values in the GGC population producing values at $12\,\rm{Gyr}$ close to 
the peak of the observed distribution.   

We should remind the readers, however, that these $r_c$ and $r_h$ values are not 
exactly the quantities observed directly.  As mentioned before, $r_c$ 
is the density-weighted core radius \citep{1985ApJ...298...80C}, related to a virial 
radius in the core, commonly used in $N$-body simulations, and can differ from an 
observed $r_c$ by a factor of a few \citep{2007MNRAS.379...93H}.  Similarly, only the 
half-light radius is observed which may differ from the half-mass radius of a cluster.  
For example,  
for a typical simulated cluster {\tt c1f3n4} the half light radius including all stars is 
$4.7\,\rm{pc}$.  If the giant stars are excluded (a common practice for observers) 
for the same cluster the half light radius is $4.1\,\rm{pc}$.  The theoretically calculated 
half mass radius for the same cluster at the same age is $7\,\rm{pc}$.  

Nevertheless, one should remember that without including stellar evolution the 
simulated $r_c/r_h$ values including 
primordial binaries were found to be about an order of magnitude smaller than in the observed 
population \citep[e.g., Paper IV, ][]{1994ApJ...431..231V} and several studies proposed different 
additional energy generation mechanisms to explain the large observed $r_c/r_h$ values 
\citep[e.g.][]{2007MNRAS.374..857T,2008IAUS..246..151C,2008ApJ...673L..25F,2008MNRAS.tmp..374M}.  
It is thus quite exciting to find such agreement simply by 
including stellar evolution in the simulations without the need for any fine tuning with the 
initial conditions or exotic scenarios.  

\begin{figure}
\begin{center}
\plotone{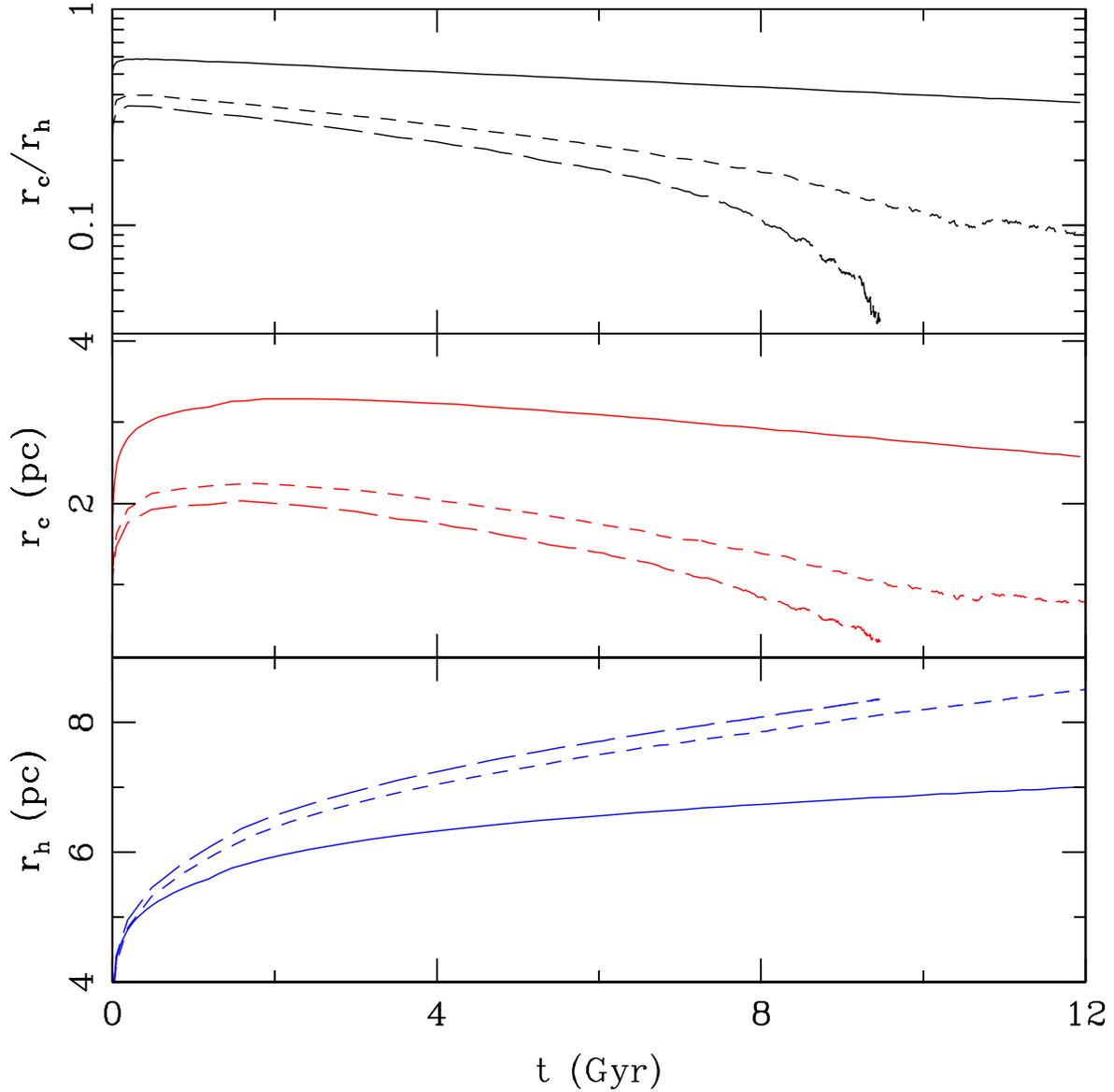}
\caption{Evolution of $r_c/r_h$ (top), $r_c$ (middle) and $r_h$ (bottom) for three 
qualitatively different clusters.  Results from runs {\tt c1f3n4}, {\tt c3f2n1}, and {\tt c8f1n1} 
are shown in all three panels with solid, short-dashed, and long-dashed lines, respectively.  
Runs {\tt c1f3n4}, {\tt c3f2n1}, and {c8f1n1} at their final stage of simulation are in the 
slow contraction, binary-burning, and deep-collapse phase, respectively.  }
\label{plot:example_rc}
\end{center}
\end{figure} 
\begin{figure}
\begin{center}
\plotone{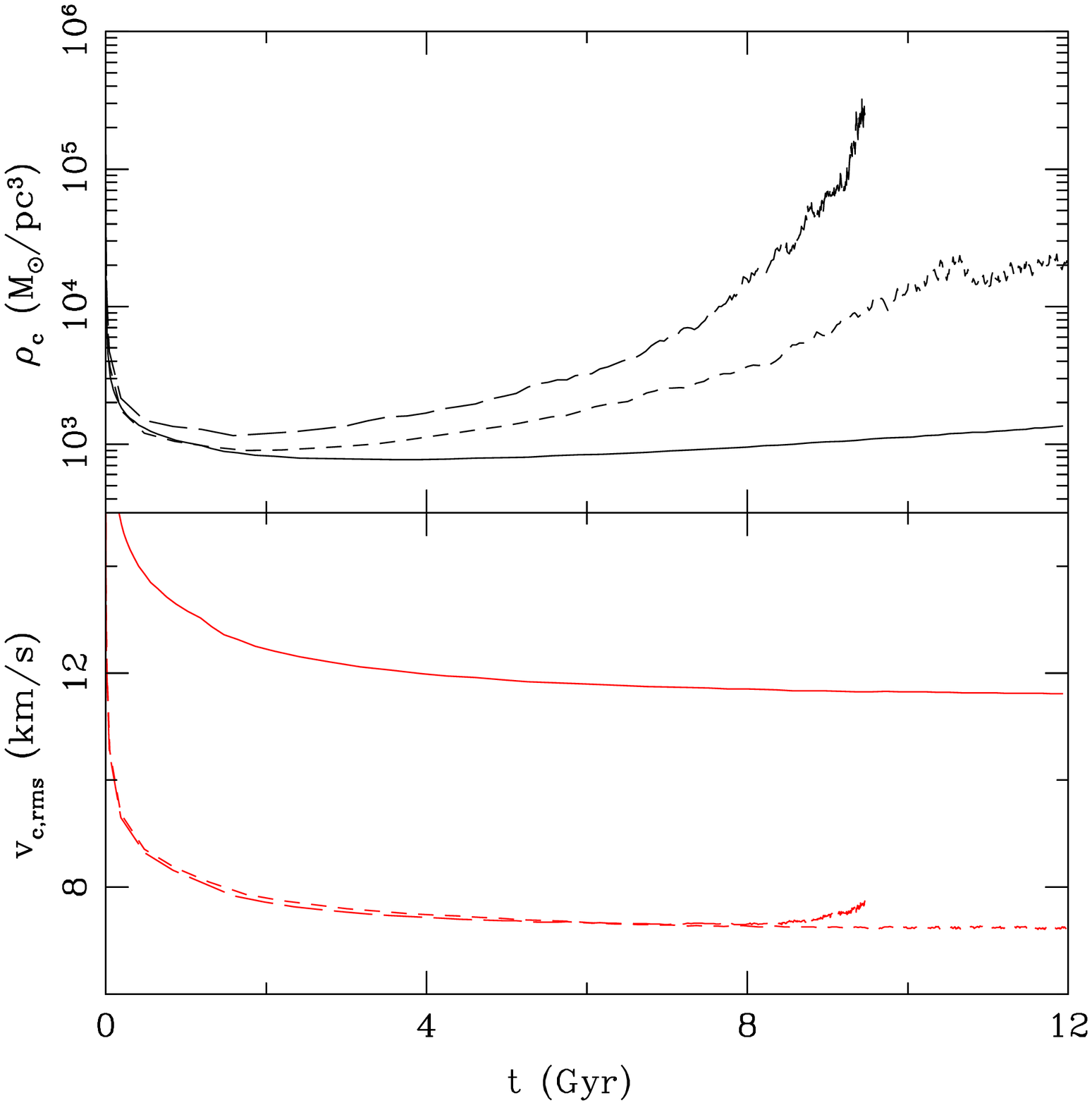}
\caption{Evolution of the central stellar mass density ($\rho_c$, top) and the central r.m.s. 
velocity ($v_{c, rms}$, bottom) for three example runs {\tt c1f3n4} (solid), {\tt c3f2n1} 
(short-dashed), and {\tt c8f1n1} (long-dashed).  $\rho_c$ and $v_{c, rms}$ decrease 
sharply during the initial stellar evolution driven expansion of the clusters.  $\rho_c$ 
increases slowly during the slow-contraction phase.  In the binary-burning phase $\rho_c$ 
attains a quasi-steady value.  A sharp increase in $\rho_c$ is observed at the deep-collapse 
phase.  $v_{c, rms}$ remains more or less steady followed by the initial decrease.  During 
deep-collapse $v_{c, rms}$ increases sharply.  }
\label{plot:example_t_rho}
\end{center}
\end{figure} 
To focus on the distinct evolutionary stages of the clusters we now choose three 
clusters from our large grid of simulations.  These clusters are representative of 
clusters in three distinct end stages.  
Cluster {\tt c1f3n4} is at the slow contraction 
phase at the integration stopping time and cluster age $t_{cl} = 12\,\rm{Gyr}$.  
Cluster {\tt c3f2n1} completes 
the slow contraction phase at $t_{cl} \sim 10\,\rm{Gyr}$, reaches the binary-burning 
quasi-steady phase and remains in the binary-burning phase until the integration stopping 
time $t_{cl} = 12\,\rm{Gyr}$.  Cluster {\tt c8f1n1} reaches the deep 
collapse phase at $t_{cl} \approx 9\,\rm{Gyr}$ (Figure\ \ref{plot:example_rc}).  Integration is 
stopped after this stage is reached.  As mentioned earlier, cluster {\tt c8f1n1} has no 
primordial binaries and shown only as a limiting case for comparison.  For each of 
the simulated clusters the three distinct phases of cluster evolution are clearly observed.  
All simulated clusters first expand due to stellar evolution mass loss during the first 
$\sim 1\,\rm{Gyr}$.  Followed by this initial expansion the clusters 
slowly contracts due to two-body relaxation.  This slow contraction phase ends in the 
quasi-steady binary-burning phase for clusters with primordial binaries 
(Table\ \ref{tab:fixedrv} runs except {\tt c$i$f1n$i$}, where $i \in [1, 4]$).  
Clusters without primordial binaries go into deep-collapse directly at the end of slow 
contraction.   

The central density for each simulated cluster first decreases sharply during the initial stellar evolution dominated phase due to the early expansion of the core.  
During the slow contraction phase the cluster stellar density 
increases steadily and reaches a quasi-steady value during the binary-burning phase 
(Figure\ \ref{plot:example_t_rho}).  The central velocity dispersion ($v_{c,rms}$) 
decreases sharply during 
the stellar evolution dominated phase.  After that $v_{c,rms}$ reaches a 
steady value of $\sim 10\,\rm{km/s}$.  The final value of $v_{c,rms}$ 
depends on the evolutionary stage of the cluster as well as the total mass in the core.  
Note that the value of $v_{c,rms}$ for runs {\tt c3f2n1} and {\tt c8f1n1} 
are similar, since the core masses are comparable, whereas a more massive cluster 
{\tt c1f3n4} shows a higher $v_{c,rms}$ as expected.  $v_{c,rms}$ for run {\tt c8f1n1} 
starts to diverge from $v_{c,rms}$ for run {\tt c3f2n1} only when the former reaches 
the deep-collapse phase.            

\begin{figure}
\begin{center}
\plotone{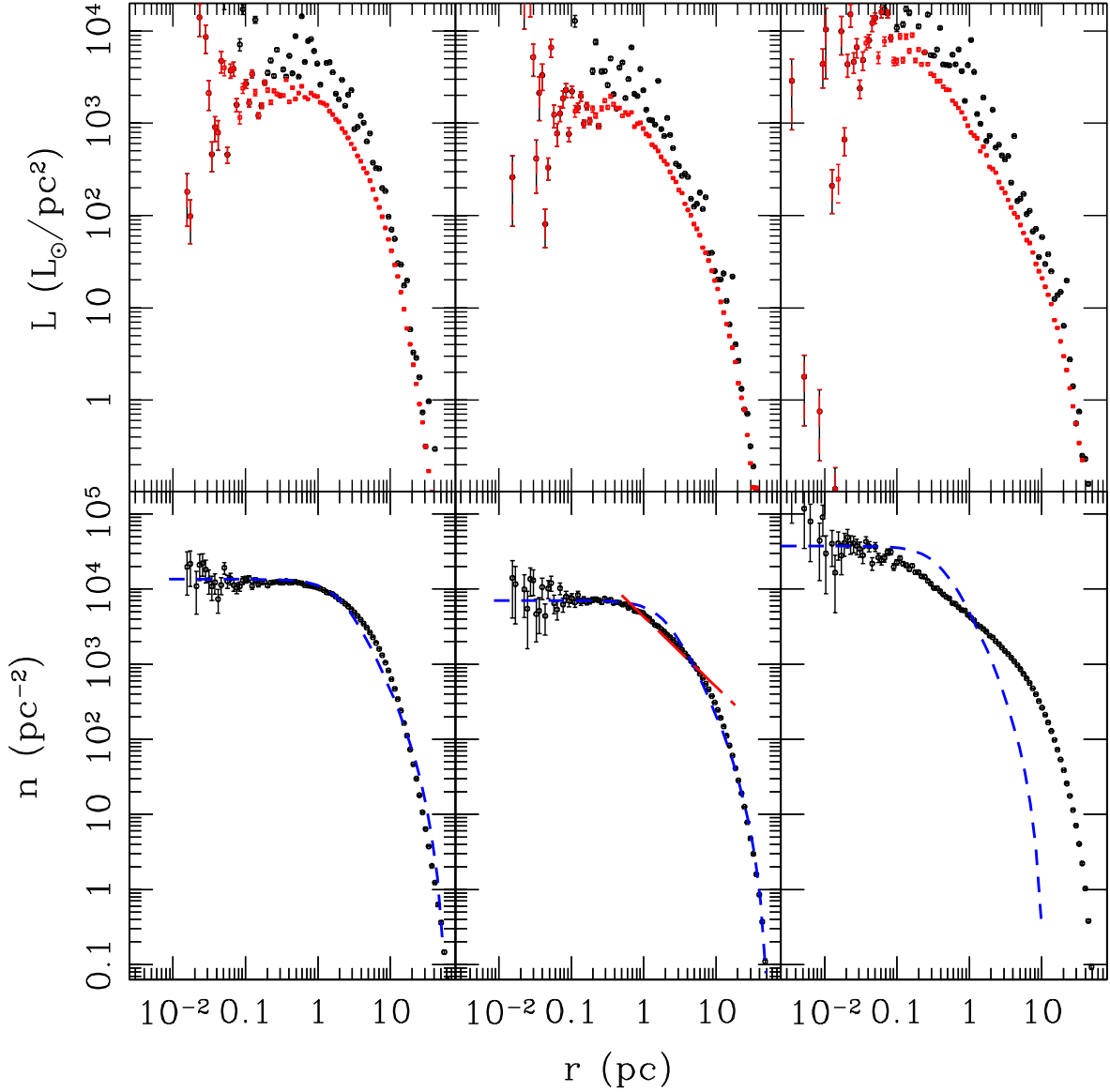}
\caption{Radial profiles for the stellar luminosity density (top) and the stellar number 
density (bottom) at $t_{cl}=12\,\rm{Gyr}$ for clusters {\tt c1f3n4}, {\tt c3f2n1}, and {c8f1n1}, 
from left to right, respectively.  The error bars on each panel show the 
Poisson noise of the data.  In the top panel the black circles show the luminosity 
density for all stars taken into account.  The red squares show the luminosity density 
taking into account only stars with stellar luminosities $L_\star < 20\,\rm{L_\odot}$.  On 
each panel the dashed blue line shows the best fit King model to the data.  Bottom 
left panel is a cluster in slow contraction phase and very well fitted by a King density 
profile.  The bottom middle panel is a cluster in the binary burning phase.  A King profile still 
works for most parts, however, a hint of a power-law profile is observed (red long dash).  
Bottom right panel is a deep-collapse cluster showing a power-law profile and a King 
profile is a very poor fit.  }
\label{plot:example_nl2d}
\end{center}
\end{figure} 
Figure\ \ref{plot:example_nl2d} shows the surface 
density profiles for the total luminosity and number of stars for 
clusters {\tt c1f3n4}, {\tt c3f2n1}, 
and {\tt c8f1n1} at the end of simulation.  For the first two clusters the $t_{cl}=12\,\rm{Gyr}$.  
The third suffers a deep-collapse at $\sim 9\,\rm{Gyr}$; the profile at the end of simulation is 
shown in that case.  
We find the best fit single-mass King profile parameters minimizing the 
$\chi^2$ statistic from a grid of detailed King models, solving the Poisson equation 
where the mass density is calculated self-consistently 
\citep[][the fitting program was kindly provided by Miocchi]{2006MNRAS.366..227M}.  
Since for old GGCs and similarly for our simulated clusters the mass range of the stars 
at the final stage is narrow, a single-mass King profile is sufficiently accurate to predict 
the cluster parameters such as the King core radius and concentration 
(see Figure\ \ref{plot:example_nl2d}).
Furthermore, we adopt a single-mass King fit since observers often follow 
this assumption \citep[e.g.,][]{2008ApJ...681..311D}.    
The deep-collapsed cluster, {\tt c8f1n1} clearly shows a very different 
projected density profile compared to the other two clusters and cannot be represented 
with a King density profile (Figure\ \ref{plot:example_nl2d}).
The collapsed cluster do not have a well defined core as seen in the steady increase in the 
stellar number surface density.  For cluster {\tt c1f3n4}, which is in the slow contraction 
phase, a standard King density profile is an excellent representation of the simulated density 
profile.  The density profile in the binary-burning cluster {\tt c3f2n1} is close to a King profile, 
however, near the central region there is a hint of a power-law density profile expected from 
observed core-collapsed clusters.  In this region, a power-law is a better representation 
than a King profile for this cluster indicating a self-similar collapse 
\citep[Figure\ \ref{plot:example_nl2d}; e.g.][]{2003gmbp.book.....H,2008gady.book.....B}.

We call the core radius and the concentrations calculated using the best fit King model 
as $r_{c,obs}$ and $c_{obs}$ respectively.  Table\ \ref{tab:fixedrv} shows a full list of these 
values for all our simulated clusters.  However, as shown in Figure\ \ref{plot:example_nl2d} 
these values are not correct for the deep-collapsed clusters.  Furthermore, for the binary burning 
clusters a King profile may not be a good fit.  
Nevertheless, most of our simulated clusters are in the slow contraction phase at 
$t_{cl} = 12\,\rm{Gyr}$, where a King density profile is an excellent fit to the data.  
The luminosity profile is noisy due to the presence of 
a few high luminosity stars (Figure\ \ref{plot:example_nl2d}).  
If only stars with a stellar luminosity $L_\star < 20\,\rm{L_\odot}$ 
are taken into account, the profile is less noisy.  

\begin{figure}
\begin{center}
\plotone{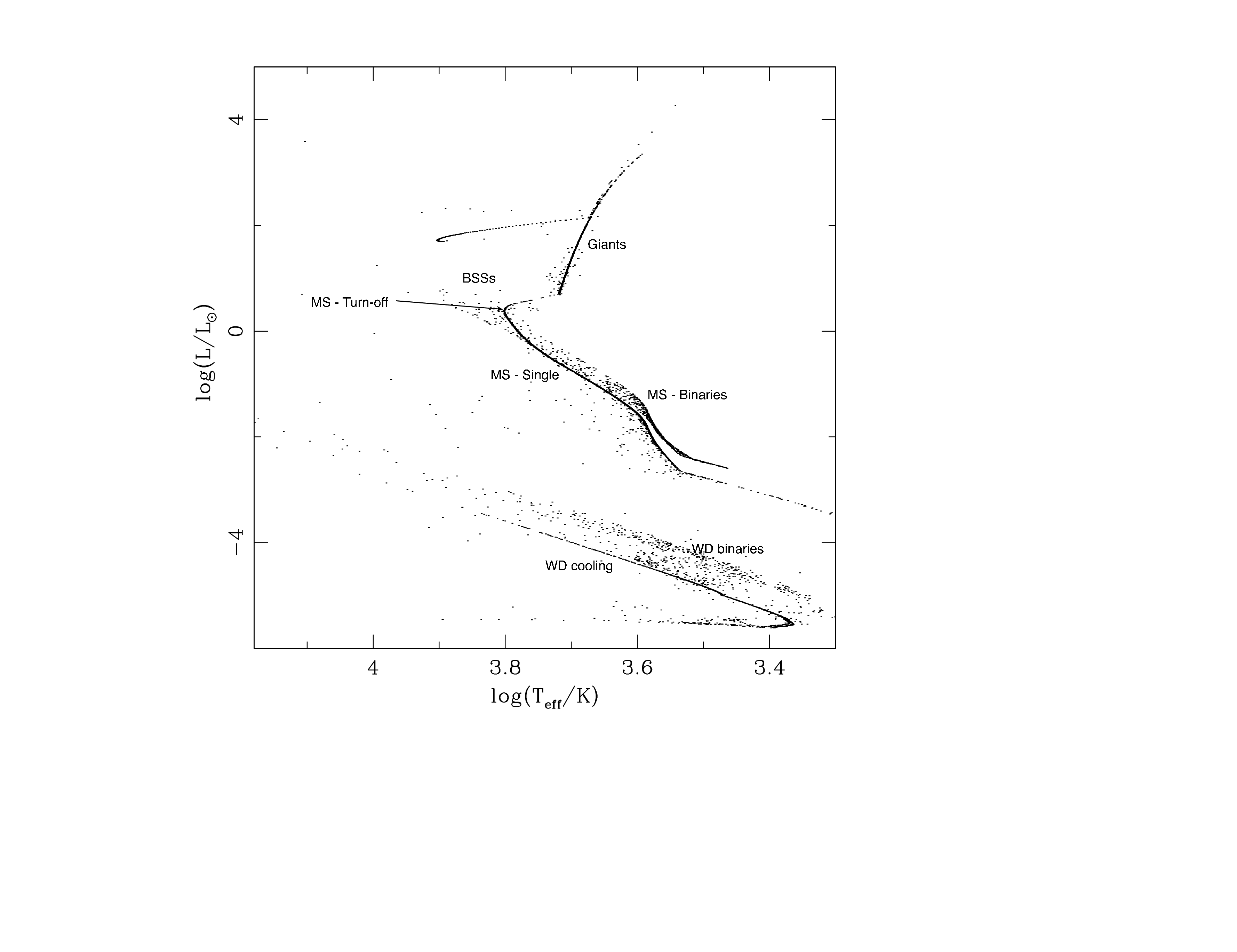}
\caption{An example of a synthetic HRD for run {\tt c1f3n4} from the set {\tt fixed-$r_v$} at 
$t_{cl}=12\,\rm{Gyr}$.  Each dot is a bound object in the cluster (a single star or a binary).  
All binaries are assumed to be unresolved.  The $T_{eff}$ for a binary is the luminosity 
weighted temperature.  The single and binary MSs of the cluster are clearly seen.  The giant 
branch, WD cooling sequence, and BSSs are also observed.  The stars in between 
the single MS and the WD binary sequence are binaries with MS and WD compact object 
components.  }
\label{plot:example_hrdiag}
\end{center}
\end{figure} 
\begin{figure}
\begin{center}
\plotone{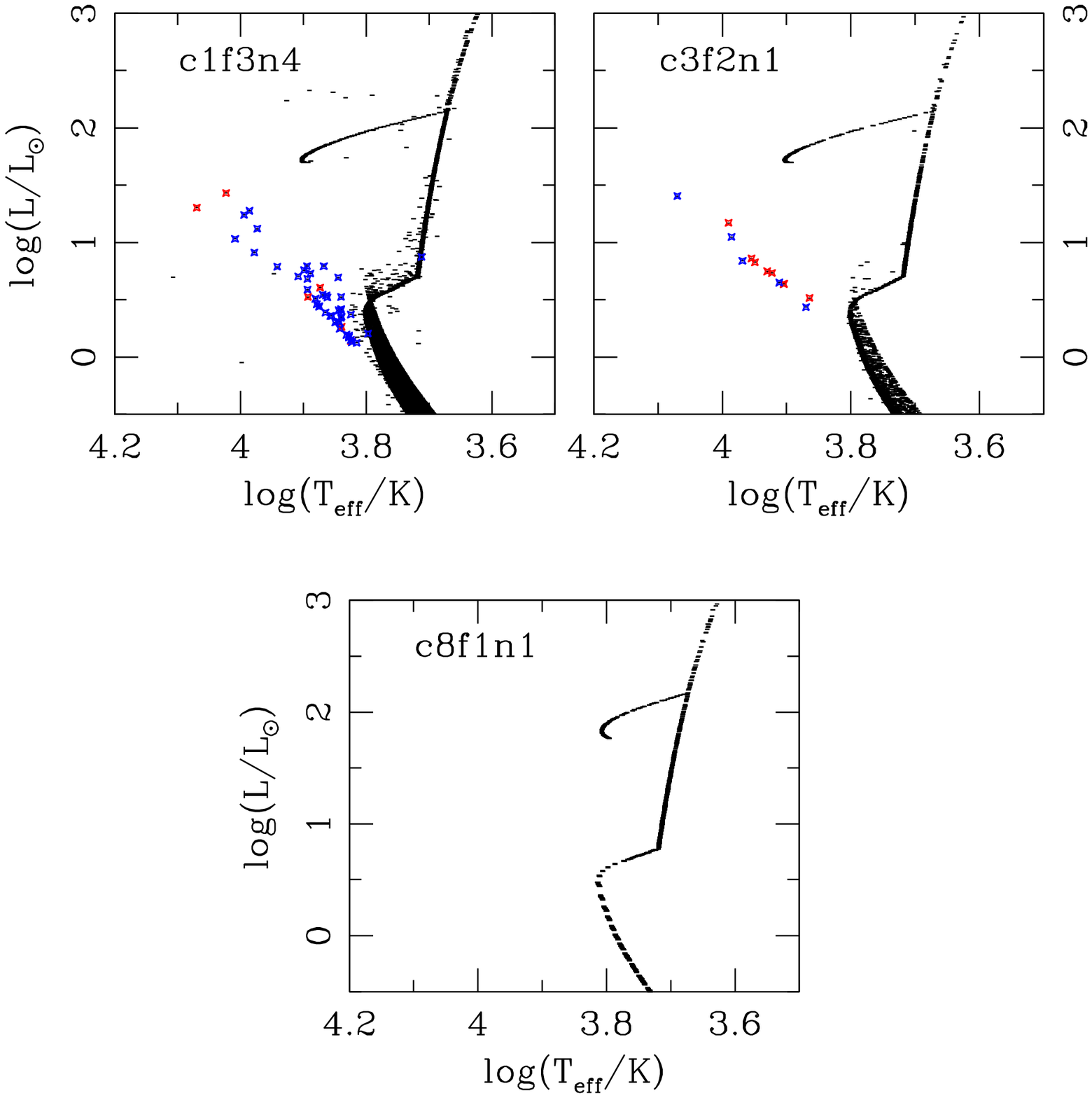}
\caption{Synthetic HRD for the three sample simulated clusters {\tt c1f3n4}, {\tt c3f2n1}, and 
{\tt c8f1n1}.  For each of the HRDs the region near the tip of the MS is shown.  
Each dot is one bound object (single or binary) in the cluster.  The BSSs 
for each cluster are shown as crosses; red and blue crosses denote 
single and binary BSSs.    Here, BSSs are defined 
as the MSSs with mass $m_\star > 1.1M_{TO}$ for the cluster at its age.  
Cluster {\tt c1f3n4} has a higher number of BSSs among the three example clusters 
($7$ singles, $45$ binaries) because of this cluster's higher $N_i$ and initial $f_b$ 
compared to cluster {\tt c3f2n1} ($9$ singles, $7$ binaries).  Cluster {\tt c8f1n1} has no 
BSSs at this age.  }
\label{plot:example_hrdiag_bss}
\end{center}
\end{figure} 
The stellar properties such as the stellar luminosity and radius for each star in the 
simulated cluster are calculated in tandem with the dynamical evolution of the cluster 
using BSE (\S\ref{method}).  
From the stellar luminosity and the radius the black-body effective temperature can 
be calculated.  Figure\ \ref{plot:example_hrdiag} shows an example of Hertzprung-Russell diagram (HRD)
obtained from the run {\tt c1f3n4} at $t_{cl} = 12\,\rm{Gyr}$.  All binaries for this cluster 
are assumed to be unresolved.  The effective temperature of a binary is approximated 
by a luminosity weighted average.  Features of a realistic HRD including the MS 
of the stars, the binary sequence, giant branch, and single and binary white dwarf 
cooling sequences can be clearly seen.  
Moreover, exotic stars such as the BSSs are 
produced.  
Here, we define any MSS with a stellar mass 
$M_\star > 1.1\,M_{TO}$ for the cluster as a BSS and find that these stars populate the 
area of HRD expected from observations (Figure\ \ref{plot:example_hrdiag_bss}).  
Here $M_{TO}$ is the MS turn-off mass 
for the cluster.  We find that the numbers of BSSs in these clusters depend on the initial 
conditions as well as the evolutionary history.  For example, clusters {\tt c1f3n4}, {\tt c3f2n1}, 
and {\tt c8f1n1} host $52$, $16$, and zero BSSs, respectively, at the time when these 
snapshots were taken ($12\,\rm{Gyr}$ for the first two and integration stopping time 
$\approx 9\,\rm{Gyr}$ for the third; Figure\ \ref{plot:example_hrdiag_bss}).  A more 
systematic study on the correlations between the total number of BSSs and the cluster 
observable properties is underway.     

\section{Summary and Conclusion}
\label{conclusion}
We report the recent update in the development of the H\'enon based MC code CMC, developed 
at Northwestern.  We have added a fitting formulae based 
single and binary stellar evolution 
using BSE by \citet{2000MNRAS.315..543H,2002MNRAS.329..897H} in addition to 
the already incorporated physical processes such as two-body relaxation and strong interaction 
including BS, BB and stellar collisions (Papers I--IV).  Thus we are now 
able to model realistic dense massive clusters 
including all relevant physics with realistic stellar IMFs in our simulations.  We test the code extensively 
and compare our results with previously published direct $N$-body 
results to validate 
CMC (\S\ref{tide},\ref{comparison}).  

In spite of the differences of the basic numerical 
methods we find that the agreement between CMC results and direct $N$-body results 
is excellent (in particular Figures\ \ref{plot:tide}-\ref{plot:hc_nbc_dd}).  
The close reproduction of the evolution of the core $f_b$ and the overall $f_b$ 
warrants special mention.  The 
evolution of $f_b$ is related to all physical processes relevant in the cluster.  Two-body 
relaxation drives mass segregation.  Binaries being more massive than typical single stars 
mass-segregate towards the center.  In the core these binaries interact, and can get 
destroyed via BS/BB interactions.  Throughout the evolution the cluster 
binaries evolve and can merge 
or disrupt simply through binary evolution.  The galactic tidal field tidally strips low mass stars 
from the cluster tidal boundary.  Thus obtaining the right evolution of $f_b$ indicates 
all these physical processes are implemented correctly. 
In addition the close agreement 
between the fraction of double-degenerate binaries in the core (Figure\ \ref{plot:hc_nbc_dd}) bolsters our belief that 
not only the dynamical processes are modeled correctly, but also stellar evolution and 
compact object formation are at least as accurate as the direct $N$-body code NBODY4.  

Although the core properties are obtained accurately using CMC, a larger difference is 
found whenever a quantity involving the total number of bound stars in the cluster 
is compared.  For example, the agreement in the evolution of the fractional number of 
bound stars, although is quite good (Figures\ \ref{plot:hc_nc}, \ref{plot:hc_nbound}) 
given the drastically 
different methods of simulations, can differ by at most $\approx 20\%$.  
These differences are dominated by the tidal mass loss effects which is 
hard to model within MC methods and can only be addressed in a criterion based 
way \citep[\S\ref{tide}; also see e.g., Paper IV][]{2000MNRAS.317..581G}.  
A more detailed study in characterizing the orbits in a cluster potential 
and tidal effects is underway, but is beyond the scope of this work.  

Our results show that including stellar evolution and a realistic IMF dramatically change 
the evolution of a star cluster (\S\ref{without}) and to model a realistic star cluster, inclusion 
of this process is vital.  The initial expansion of the cluster driven by stellar evolution mass 
loss significantly prolongs the slow contraction phase.  Depending on the initial properties 
of the cluster, even without any primordial binaries the slow contraction phase may last 
more than a Hubble time for clusters typical for the GGCs 
(e.g., Figures\ \ref{plot:senose}, \ref{plot:rv_rc_phys}).  
On the contrary, exclusion of this important effect leads to a quick contraction of the 
cluster due to mass segregation even if only a moderately broad range of stellar IMF 
is used (Figure\ \ref{plot:senose}).
We also show that even for 
simulations with a very narrow stellar mass range, for a relatively low $f_b$ 
(Figure\ \ref{plot:w4n1e5}) the inclusion of stellar evolution can 
give $r_c/r_h$ values $\approx 10\%$ larger compared to when it is left out.       

One of the biggest uncertainties of studying the evolution of dense, massive star clusters 
is in determining the initial conditions.  The detailed evolution of a cluster depends on 
various initial properties 
including the initial effective radius, mass, $f_b$, concentration, and the galactic tidal field 
and estimating the initial conditions using the present day observed properties of a cluster 
is hard.  Moreover, starting from different initial conditions it is possible to achieve very 
similar present day properties, e.g., $r_c$, and $r_c/r_h$ 
(Figures\ \ref{plot:rv_rc_phys} and \ref{plot:rv_rcoverrh}).  
In addition, the observed present 
day values can also be quite uncertain, especially the 3D orbit of a cluster in the galactic 
potential is hard to measure.  Furthermore, although it may be possible to qualitatively understand 
individual effects of the various physical processes on the observable cluster properties, 
the collective effect is impossible to judge without actual detailed simulations.      
Thus to understand a population of dense clusters it is required to study a large parameter 
space and study evolution of these cluster in a realistic way including all physics in tandem.  

With the recent improvement to CMC it is 
now possible to truly scan the full parameter space realistically without any loss of 
generality due to its significantly lower computational cost compared to the direct 
$N$-body codes and the accuracy and ability to treat all relevant physical processes.  
We have started a detailed study to create a population of realistic globular clusters, 
representative of the observed GGCs with a large grid of simulations with realistic initial 
conditions motivated by observations of young massive clusters 
\citep[e.g.][]{2007A&A...469..925S,2009gcgg.book..103S}.  
Here we have presented some of these simulations to show that rather than 
creating specific clusters it is now beginning to be possible to create a whole population 
of GGCs using CMC in a star-by-star detail.  
We show that using observationally motivated initial conditions, without any need for 
fine tuning it is possible to create old dense clusters very similar to the observed GGCs 
(Figures\ \ref{plot:rv_rc_phys} -- \ref{plot:rv_rcoverrh}).  

Each star in CMC has realistic stellar properties 
such as luminosity, radius, 
and effective temperature in addition to the mass and position in the cluster (which is 
sufficient to follow its dynamics).  Hence, in addition to the global evolution of the clusters 
it is possible to study individual stellar populations in a cluster.  For example, we show 
synthetic HRDs for a few simulated clusters from our grid of simulations.  All features, 
including, e.g., the single and binary MS, WD cooling sequence, the giant branch, 
and BSSs, of a realistic HRD can be seen in the synthetic HRD 
(Figures\ \ref{plot:example_hrdiag}, \ref{plot:example_hrdiag_bss}).  
After this crucial improvement to CMC a large array of interesting problems are now 
accessible.   
For example, a detailed study of the observed GGC BSSs, their 
properties, and the correlations with various cluster properties is 
underway.  

\begin{deluxetable}{cccccccccc|cccccccccc}
\tabletypesize{\footnotesize}
\rotate
\tablecolumns{20}
\tablewidth{0pt}
\tablecaption{List of simulations: $W_0$ is the central concentration parameter for a King 
profile \citep{1966AJ.....71...64K}, 
cluster mass $M$ is in $10^5\,\rm{M_\odot}$, 
number of bound cluster objects $N$ is given in $10^5$,   
central stellar mass density $\rho_c$ is in $10^3\,\rm{M_\odot/pc^3}$, 
$r_c$ and $r_h$ are in pc, and 
$c$ is the concentration parameter defined as $log_{10}(r_c/r_t)$.   
All final values are extracted from the final snapshot of the simulated clusters.
$r_{c, obs}$ and $c_{obs}$ are estimated from a single-mass best fit King model 
to the $2D$ number density at the final snapshot of the cluster.    
\label{tab:fixedrv}}
\tablehead{
\colhead{Name} & 
\multicolumn{9}{c}{Initial} & 
\multicolumn{10}{c}{Final} \\ 
\hline\\
\colhead{} & 
\colhead{$W_0$} & 
\colhead{$M$} & 
\colhead{$N$} & 
\colhead{$r_c$} & 
\colhead{$r_h$} & 
\colhead{$\rho_c$} & 
\colhead{$f_b$} & 
\colhead{$f_{b,c}$} & 
\colhead{$c$} & 
\colhead{$M$} & 
\colhead{$N$} & 
\colhead{$r_c$} & 
\colhead{$r_{c,obs}$} & 
\colhead{$r_h$} & 
\colhead{$\rho_c$} & 
\colhead{$f_b$} & 
\colhead{$f_{b,c}$} & 
\colhead{$c$} & 
\colhead{$c_{obs}$} } 
\startdata
c1f1n1 & 4 & 2.5 & 4 & 1.6 & 3.3 & 12.2 & 0.00 & 0.00 & 1.1 & 1.4 & 3 & 2.0 & 2.1 & 7.1 & 1.0 & 0.00 & 0.00 & 1.5 & 1.5 \\ 
c1f1n2 & 4 & 3.8 & 6 & 1.6 & 3.3 & 16.2 & 0.00 & 0.00 & 1.5 & 2.1 & 5 & 2.3 & 2.2 & 7.0 & 1.1 & 0.00 & 0.00 & 1.5 & 1.5 \\ 
c1f1n3 & 4 & 5.1 & 8 & 1.6 & 3.3 & 23.3 & 0.00 & 0.00 & 1.7 & 2.8 & 7 & 2.4 & 2.1 & 6.9 & 1.2 & 0.00 & 0.00 & 1.5 & 1.5 \\ 
c1f1n4 & 4 & 6.4 & 10 & 1.6 & 3.3 & 29.0 & 0.00 & 0.00 & 1.6 & 3.5 & 9 & 2.5 & 2.0 & 6.8 & 1.4 & 0.00 & 0.00 & 1.5 & 1.6 \\ 
c1f2n1 & 4 & 2.6 & 4 & 1.6 & 3.3 & 12.3 & 0.05 & 0.05 & 1.3 & 1.4 & 3 & 2.1 & 2.3 & 7.2 & 0.9 & 0.05 & 0.07 & 1.4 & 1.5 \\ 
c1f2n2 & 4 & 3.9 & 6 & 1.6 & 3.3 & 17.3 & 0.05 & 0.05 & 1.6 & 2.1 & 5 & 2.3 & 2.2 & 7.1 & 1.1 & 0.05 & 0.07 & 1.5 & 1.5 \\ 
c1f2n3 & 4 & 5.3 & 8 & 1.6 & 3.3 & 24.7 & 0.05 & 0.05 & 1.7 & 2.9 & 7 & 2.5 & 2.2 & 7.0 & 1.2 & 0.05 & 0.07 & 1.5 & 1.5 \\ 
c1f2n4 & 4 & 6.6 & 10 & 1.6 & 3.3 & 30.2 & 0.05 & 0.05 & 1.7 & 3.6 & 9 & 2.5 & 2.0 & 6.9 & 1.4 & 0.05 & 0.06 & 1.5 & 1.6 \\ 
c1f3n1 & 4 & 2.7 & 4 & 1.6 & 3.3 & 12.4 & 0.10 & 0.10 & 1.2 & 1.4 & 3 & 2.1 & 2.3 & 7.3 & 0.9 & 0.09 & 0.14 & 1.4 & 1.5 \\ 
c1f3n2 & 4 & 4.0 & 6 & 1.6 & 3.3 & 17.7 & 0.10 & 0.10 & 1.3 & 2.2 & 5 & 2.3 & 2.2 & 7.2 & 1.1 & 0.09 & 0.14 & 1.5 & 1.5 \\ 
c1f3n3 & 4 & 5.4 & 8 & 1.6 & 3.3 & 25.3 & 0.10 & 0.10 & 1.5 & 2.9 & 7 & 2.5 & 2.2 & 7.1 & 1.2 & 0.09 & 0.13 & 1.5 & 1.5 \\ 
c1f3n4 & 4 & 6.8 & 10 & 1.6 & 3.3 & 30.9 & 0.10 & 0.10 & 1.7 & 3.7 & 9 & 2.6 & 2.1 & 7.0 & 1.4 & 0.09 & 0.12 & 1.5 & 1.5 \\ 
c2f1n1 & 4.5 & 2.5 & 4 & 1.5 & 3.3 & 14.2 & 0.00 & 0.00 & 1.0 & 1.4 & 3 & 1.9 & 2.0 & 7.1 & 1.3 & 0.00 & 0.00 & 1.5 & 1.5 \\ 
c2f1n2 & 4.5 & 3.8 & 6 & 1.5 & 3.3 & 18.8 & 0.00 & 0.00 & 1.5 & 2.1 & 5 & 2.1 & 2.0 & 7.0 & 1.4 & 0.00 & 0.00 & 1.5 & 1.5 \\ 
c2f1n3 & 4.5 & 5.1 & 8 & 1.5 & 3.3 & 27.2 & 0.00 & 0.00 & 1.4 & 2.8 & 7 & 2.3 & 2.0 & 6.9 & 1.5 & 0.00 & 0.00 & 1.5 & 1.6 \\ 
c2f1n4 & 4.5 & 6.4 & 10 & 1.5 & 3.3 & 33.3 & 0.00 & 0.00 & 1.4 & 3.5 & 9 & 2.3 & 2.0 & 6.9 & 1.7 & 0.00 & 0.00 & 1.5 & 1.6 \\ 
c2f2n1 & 4.5 & 2.6 & 4 & 1.5 & 3.3 & 14.3 & 0.05 & 0.05 & 1.0 & 1.4 & 3 & 1.9 & 2.1 & 7.3 & 1.2 & 0.05 & 0.07 & 1.5 & 1.5 \\ 
c2f2n2 & 4.5 & 3.9 & 6 & 1.5 & 3.3 & 20.1 & 0.05 & 0.05 & 1.7 & 2.1 & 5 & 2.2 & 2.1 & 7.2 & 1.3 & 0.05 & 0.07 & 1.5 & 1.5 \\ 
c2f2n3 & 4.5 & 5.3 & 8 & 1.5 & 3.3 & 28.9 & 0.05 & 0.05 & 1.6 & 2.8 & 7 & 2.3 & 2.1 & 7.1 & 1.4 & 0.05 & 0.07 & 1.5 & 1.5 \\ 
c2f2n4 & 4.5 & 6.6 & 10 & 1.5 & 3.3 & 34.7 & 0.05 & 0.05 & 1.7 & 3.6 & 9 & 2.4 & 2.0 & 7.0 & 1.7 & 0.05 & 0.07 & 1.5 & 1.6 \\ 
c2f3n1 & 4.5 & 2.7 & 4 & 1.5 & 3.3 & 14.4 & 0.10 & 0.10 & 1.0 & 1.4 & 3 & 2.0 & 2.2 & 7.4 & 1.2 & 0.09 & 0.14 & 1.5 & 1.5 \\ 
c2f3n2 & 4.5 & 4.0 & 6 & 1.5 & 3.3 & 20.5 & 0.10 & 0.10 & 1.8 & 2.2 & 5 & 2.2 & 2.1 & 7.3 & 1.3 & 0.09 & 0.14 & 1.5 & 1.5 \\ 
c2f3n3 & 4.5 & 5.4 & 8 & 1.5 & 3.3 & 29.6 & 0.10 & 0.10 & 1.4 & 2.9 & 7 & 2.4 & 2.1 & 7.2 & 1.3 & 0.09 & 0.13 & 1.5 & 1.5 \\ 
c2f3n4 & 4.5 & 6.8 & 10 & 1.5 & 3.3 & 35.4 & 0.10 & 0.10 & 1.4 & 3.7 & 9 & 2.4 & 2.0 & 7.1 & 1.7 & 0.09 & 0.13 & 1.5 & 1.6 \\ 
c3f1n1 & 5 & 2.5 & 4 & 1.4 & 3.2 & 17.2 & 0.00 & 0.00 & 1.2 & 1.4 & 3 & 1.7 & 1.9 & 7.3 & 1.7 & 0.00 & 0.00 & 1.5 & 1.5 \\ 
c3f1n2 & 5 & 3.8 & 6 & 1.4 & 3.3 & 22.4 & 0.00 & 0.00 & 1.4 & 2.1 & 5 & 2.0 & 2.0 & 7.1 & 1.7 & 0.00 & 0.00 & 1.5 & 1.6 \\ 
c3f1n3 & 5 & 5.1 & 8 & 1.4 & 3.3 & 32.5 & 0.00 & 0.00 & 1.7 & 2.8 & 7 & 2.1 & 2.0 & 7.0 & 1.9 & 0.00 & 0.00 & 1.6 & 1.6 \\ 
c3f1n4 & 5 & 6.4 & 10 & 1.4 & 3.3 & 40.0 & 0.00 & 0.00 & 1.6 & 3.5 & 9 & 2.2 & 2.0 & 7.0 & 2.1 & 0.00 & 0.00 & 1.6 & 1.6 \\ 
c3f2n1 & 5 & 2.6 & 4 & 1.4 & 3.2 & 17.3 & 0.05 & 0.05 & 1.3 & 1.4 & 3 & 1.7 & 2.1 & 7.4 & 1.7 & 0.05 & 0.08 & 1.5 & 1.5 \\ 
c3f2n2 & 5 & 3.9 & 6 & 1.4 & 3.3 & 24.0 & 0.05 & 0.05 & 1.3 & 2.1 & 5 & 2.0 & 2.0 & 7.3 & 1.6 & 0.05 & 0.07 & 1.5 & 1.5 \\ 
c3f2n3 & 5 & 5.3 & 8 & 1.4 & 3.3 & 34.6 & 0.05 & 0.05 & 1.5 & 2.8 & 7 & 2.2 & 2.0 & 7.2 & 1.8 & 0.05 & 0.07 & 1.5 & 1.6 \\ 
c3f2n4 & 5 & 6.6 & 10 & 1.4 & 3.3 & 41.6 & 0.05 & 0.05 & 1.7 & 3.6 & 9 & 2.2 & 2.0 & 7.1 & 2.0 & 0.05 & 0.07 & 1.5 & 1.6 \\ 
c3f3n1 & 5 & 2.7 & 4 & 1.4 & 3.2 & 17.4 & 0.10 & 0.10 & 1.5 & 1.4 & 3 & 1.8 & 2.1 & 7.6 & 1.5 & 0.09 & 0.15 & 1.5 & 1.5 \\ 
c3f3n2 & 5 & 4.0 & 6 & 1.4 & 3.3 & 24.5 & 0.10 & 0.10 & 1.5 & 2.2 & 5 & 2.0 & 2.1 & 7.4 & 1.6 & 0.09 & 0.14 & 1.5 & 1.5 \\ 
c3f3n3 & 5 & 5.4 & 8 & 1.4 & 3.3 & 35.4 & 0.10 & 0.10 & 1.4 & 2.9 & 7 & 2.2 & 2.1 & 7.3 & 1.8 & 0.09 & 0.13 & 1.5 & 1.6 \\ 
c3f3n4 & 5 & 6.8 & 10 & 1.4 & 3.3 & 42.5 & 0.10 & 0.10 & 1.5 & 3.6 & 9 & 2.3 & 2.0 & 7.2 & 2.0 & 0.09 & 0.13 & 1.5 & 1.6 \\ 
c4f1n1 & 5.5 & 2.5 & 4 & 1.3 & 3.2 & 21.6 & 0.00 & 0.00 & 1.2 & 1.4 & 3 & 1.5 & 1.8 & 7.4 & 2.5 & 0.00 & 0.00 & 1.6 & 1.6 \\ 
c4f1n2 & 5.5 & 3.8 & 6 & 1.3 & 3.2 & 27.6 & 0.00 & 0.00 & 1.7 & 2.1 & 5 & 1.8 & 1.9 & 7.3 & 2.4 & 0.00 & 0.00 & 1.6 & 1.6 \\ 
c4f1n3 & 5.5 & 5.1 & 8 & 1.3 & 3.3 & 40.2 & 0.00 & 0.00 & 1.4 & 2.8 & 7 & 1.8 & 2.0 & 7.2 & 2.9 & 0.00 & 0.00 & 1.6 & 1.6 \\ 
c4f1n4 & 5.5 & 6.4 & 10 & 1.3 & 3.2 & 49.6 & 0.00 & 0.00 & 1.8 & 3.5 & 9 & 2.0 & 2.0 & 7.1 & 2.9 & 0.00 & 0.00 & 1.6 & 1.6 \\ 
c4f2n1 & 5.5 & 2.6 & 4 & 1.3 & 3.2 & 21.7 & 0.05 & 0.05 & 1.6 & 1.4 & 3 & 1.6 & 1.9 & 7.6 & 2.1 & 0.05 & 0.08 & 1.5 & 1.6 \\ 
c4f2n2 & 5.5 & 3.9 & 6 & 1.3 & 3.2 & 29.7 & 0.05 & 0.05 & 1.3 & 2.1 & 5 & 1.8 & 1.9 & 7.4 & 2.4 & 0.05 & 0.08 & 1.6 & 1.6 \\ 
c4f2n3 & 5.5 & 5.3 & 8 & 1.3 & 3.3 & 43.0 & 0.05 & 0.05 & 1.6 & 2.8 & 7 & 1.9 & 2.0 & 7.3 & 2.6 & 0.05 & 0.08 & 1.6 & 1.6 \\ 
c4f2n4 & 5.5 & 6.6 & 10 & 1.3 & 3.2 & 51.5 & 0.05 & 0.05 & 1.8 & 3.6 & 9 & 2.0 & 2.0 & 7.2 & 2.8 & 0.05 & 0.07 & 1.6 & 1.6 \\ 
c4f3n1 & 5.5 & 2.7 & 4 & 1.3 & 3.2 & 21.7 & 0.10 & 0.10 & 1.3 & 1.4 & 3 & 1.6 & 1.9 & 7.7 & 2.0 & 0.09 & 0.15 & 1.5 & 1.6 \\ 
c4f3n2 & 5.5 & 4.0 & 6 & 1.3 & 3.2 & 30.3 & 0.10 & 0.10 & 1.3 & 2.2 & 5 & 1.9 & 1.9 & 7.5 & 2.3 & 0.09 & 0.15 & 1.6 & 1.6 \\ 
c4f3n3 & 5.5 & 5.4 & 8 & 1.3 & 3.3 & 44.0 & 0.10 & 0.10 & 1.3 & 2.9 & 7 & 2.0 & 2.0 & 7.4 & 2.5 & 0.09 & 0.14 & 1.6 & 1.6 \\ 
c4f3n4 & 5.5 & 6.8 & 10 & 1.3 & 3.2 & 52.5 & 0.10 & 0.10 & 1.8 & 3.6 & 9 & 2.1 & 2.0 & 7.3 & 2.6 & 0.09 & 0.13 & 1.6 & 1.6 \\ 
c5f1n1 & 6 & 2.5 & 4 & 1.2 & 3.2 & 28.5 & 0.00 & 0.00 & 1.4 & 1.4 & 3 & 1.2 & 1.4 & 7.6 & 4.9 & 0.00 & 0.00 & 1.7 & 1.7 \\ 
c5f1n2 & 6 & 3.8 & 6 & 1.2 & 3.2 & 35.7 & 0.00 & 0.00 & 1.3 & 2.1 & 5 & 1.4 & 1.4 & 7.5 & 4.7 & 0.00 & 0.00 & 1.7 & 1.8 \\ 
c5f1n3 & 6 & 5.1 & 8 & 1.2 & 3.2 & 52.8 & 0.00 & 0.00 & 1.4 & 2.8 & 7 & 1.6 & 1.9 & 7.4 & 4.5 & 0.00 & 0.00 & 1.7 & 1.6 \\ 
c5f1n4 & 6 & 6.4 & 10 & 1.2 & 3.2 & 64.3 & 0.00 & 0.00 & 1.7 & 3.5 & 9 & 1.7 & 2.0 & 7.3 & 4.9 & 0.00 & 0.00 & 1.7 & 1.6 \\ 
c5f2n1 & 6 & 2.6 & 4 & 1.2 & 3.2 & 28.6 & 0.05 & 0.05 & 1.4 & 1.4 & 3 & 1.3 & 1.4 & 7.8 & 3.8 & 0.05 & 0.09 & 1.6 & 1.7 \\ 
c5f2n2 & 6 & 3.9 & 6 & 1.2 & 3.2 & 38.5 & 0.05 & 0.05 & 1.3 & 2.1 & 5 & 1.5 & 1.9 & 7.7 & 4.1 & 0.05 & 0.08 & 1.6 & 1.6 \\ 
c5f2n3 & 6 & 5.3 & 8 & 1.2 & 3.2 & 56.6 & 0.05 & 0.05 & 1.3 & 2.8 & 7 & 1.7 & 2.0 & 7.5 & 4.0 & 0.05 & 0.08 & 1.6 & 1.6 \\ 
c5f2n4 & 6 & 6.6 & 10 & 1.2 & 3.2 & 66.5 & 0.05 & 0.05 & 1.7 & 3.6 & 9 & 1.8 & 2.0 & 7.4 & 4.3 & 0.05 & 0.07 & 1.6 & 1.6 \\ 
c5f3n1 & 6 & 2.7 & 4 & 1.2 & 3.2 & 28.6 & 0.10 & 0.10 & 1.5 & 1.4 & 3 & 1.4 & 1.7 & 8.0 & 3.5 & 0.09 & 0.16 & 1.6 & 1.7 \\ 
c5f3n2 & 6 & 4.0 & 6 & 1.2 & 3.2 & 39.3 & 0.10 & 0.10 & 1.3 & 2.2 & 5 & 1.6 & 1.9 & 7.8 & 3.8 & 0.09 & 0.16 & 1.6 & 1.6 \\ 
c5f3n3 & 6 & 5.4 & 8 & 1.2 & 3.2 & 57.9 & 0.10 & 0.10 & 1.4 & 2.9 & 7 & 1.7 & 2.0 & 7.6 & 3.9 & 0.09 & 0.14 & 1.6 & 1.6 \\ 
c5f3n4 & 6 & 6.8 & 10 & 1.2 & 3.2 & 67.9 & 0.10 & 0.10 & 1.8 & 3.8 & 9 & 2.3 & 1.9 & 6.9 & 1.9 & 0.09 & 0.13 & 1.5 & 1.6 \\ 
c6f1n1 & 6.5 & 2.5 & 4 & 1.1 & 3.2 & 40.1 & 0.00 & 0.00 & 1.3 & 1.4 & 3 & 0.6 & 0.6 & 8.0 & 47.6 & 0.00 & 0.00 & 2.0 & 2.1 \\ 
c6f1n2 & 6.5 & 3.8 & 6 & 1.1 & 3.2 & 49.3 & 0.00 & 0.00 & 1.4 & 2.1 & 5 & 1.1 & 1.2 & 7.7 & 9.7 & 0.00 & 0.00 & 1.8 & 1.8 \\ 
c6f1n3 & 6.5 & 5.1 & 8 & 1.1 & 3.2 & 73.0 & 0.00 & 0.00 & 1.7 & 2.8 & 7 & 1.2 & 1.3 & 7.6 & 11.6 & 0.00 & 0.00 & 1.8 & 1.8 \\ 
c6f1n4 & 6.5 & 6.4 & 10 & 1.1 & 3.2 & 88.8 & 0.00 & 0.00 & 1.6 & 3.5 & 9 & 1.3 & 1.3 & 7.5 & 9.4 & 0.00 & 0.00 & 1.8 & 1.8 \\ 
c6f2n1 & 6.5 & 2.6 & 4 & 1.1 & 3.2 & 40.3 & 0.05 & 0.05 & 1.5 & 1.4 & 3 & 0.7 & 1.0 & 8.1 & 25.6 & 0.05 & 0.11 & 1.9 & 1.9 \\ 
c6f2n2 & 6.5 & 3.9 & 6 & 1.1 & 3.2 & 53.4 & 0.05 & 0.05 & 1.4 & 2.1 & 5 & 1.1 & 1.3 & 7.9 & 9.6 & 0.05 & 0.09 & 1.7 & 1.8 \\ 
c6f2n3 & 6.5 & 5.3 & 8 & 1.1 & 3.2 & 78.5 & 0.05 & 0.05 & 1.7 & 2.8 & 7 & 1.3 & 1.4 & 7.8 & 8.1 & 0.05 & 0.08 & 1.7 & 1.8 \\ 
c6f2n4 & 6.5 & 6.6 & 10 & 1.1 & 3.2 & 91.7 & 0.05 & 0.05 & 1.6 & 3.6 & 9 & 1.4 & 1.4 & 7.7 & 8.7 & 0.05 & 0.08 & 1.7 & 1.8 \\ 
c6f3n1 & 6.5 & 2.7 & 4 & 1.1 & 3.2 & 40.2 & 0.10 & 0.10 & 1.4 & 1.4 & 3 & 1.0 & 1.3 & 8.2 & 11.2 & 0.09 & 0.18 & 1.8 & 1.8 \\ 
c6f3n2 & 6.5 & 4.0 & 6 & 1.1 & 3.2 & 54.4 & 0.10 & 0.10 & 1.4 & 2.2 & 5 & 1.2 & 1.5 & 8.0 & 8.3 & 0.09 & 0.16 & 1.7 & 1.8 \\ 
c6f3n3 & 6.5 & 5.4 & 8 & 1.1 & 3.2 & 80.2 & 0.10 & 0.10 & 1.4 & 2.9 & 7 & 1.4 & 1.5 & 7.9 & 8.1 & 0.09 & 0.15 & 1.7 & 1.8 \\ 
c6f3n4 & 6.5 & 6.8 & 10 & 1.1 & 3.2 & 93.6 & 0.10 & 0.10 & 1.6 & 3.6 & 9 & 1.4 & 1.4 & 7.8 & 9.4 & 0.09 & 0.15 & 1.7 & 1.8 \\ 
c7f1n1 & 7 & 2.5 & 4 & 0.9 & 3.2 & 61.0 & 0.00 & 0.00 & 1.5 & 1.4 & 3 & 0.3 & 1.3 & 8.2 & 241.9 & 0.00 & 0.00 & 2.3 & 1.8 \\ 
c7f1n2 & 7 & 3.8 & 6 & 0.9 & 3.3 & 73.4 & 0.00 & 0.00 & 1.5 & 2.1 & 5 & 0.3 & 1.0 & 8.2 & 314.0 & 0.00 & 0.00 & 2.3 & 1.9 \\ 
c7f1n3 & 7 & 5.1 & 8 & 0.9 & 3.3 & 110.3 & 0.00 & 0.00 & 1.5 & 2.8 & 7 & 0.7 & 0.8 & 8.0 & 56.3 & 0.00 & 0.00 & 2.0 & 2.1 \\ 
c7f1n4 & 7 & 6.4 & 10 & 0.9 & 3.3 & 134.5 & 0.00 & 0.00 & 1.6 & 3.5 & 9 & 0.8 & 0.8 & 7.9 & 54.2 & 0.00 & 0.00 & 2.0 & 2.1 \\ 
c7f2n1 & 7 & 2.6 & 4 & 0.9 & 3.2 & 61.1 & 0.05 & 0.05 & 1.5 & 1.4 & 3 & 0.8 & 0.8 & 8.5 & 19.3 & 0.05 & 0.11 & 1.8 & 2.0 \\ 
c7f2n2 & 7 & 3.9 & 6 & 0.9 & 3.3 & 79.7 & 0.05 & 0.05 & 1.9 & 2.1 & 5 & 0.7 & 0.9 & 8.3 & 43.3 & 0.05 & 0.11 & 1.9 & 2.0 \\ 
c7f2n3 & 7 & 5.3 & 8 & 0.9 & 3.3 & 118.9 & 0.05 & 0.05 & 1.5 & 2.8 & 7 & 0.9 & 1.1 & 8.2 & 29.5 & 0.05 & 0.10 & 1.9 & 1.9 \\ 
c7f2n4 & 7 & 6.6 & 10 & 0.9 & 3.3 & 138.3 & 0.05 & 0.05 & 1.6 & 3.6 & 9 & 0.8 & 0.8 & 8.0 & 48.4 & 0.04 & 0.09 & 2.0 & 2.1 \\ 
c7f3n1 & 7 & 2.7 & 4 & 0.9 & 3.2 & 60.8 & 0.10 & 0.10 & 1.7 & 1.4 & 3 & 0.9 & 1.1 & 8.6 & 17.4 & 0.09 & 0.19 & 1.8 & 1.8 \\ 
c7f3n2 & 7 & 4.0 & 6 & 0.9 & 3.3 & 81.2 & 0.10 & 0.10 & 1.6 & 2.2 & 5 & 0.8 & 1.0 & 8.5 & 30.1 & 0.09 & 0.19 & 1.9 & 2.0 \\ 
c7f3n3 & 7 & 5.4 & 8 & 0.9 & 3.3 & 121.3 & 0.10 & 0.10 & 1.5 & 2.9 & 7 & 1.1 & 1.2 & 8.2 & 15.9 & 0.09 & 0.17 & 1.8 & 1.8 \\ 
c7f3n4 & 7 & 6.8 & 10 & 0.9 & 3.3 & 141.0 & 0.10 & 0.10 & 1.9 & 3.6 & 9 & 1.0 & 1.2 & 8.1 & 25.5 & 0.09 & 0.16 & 1.9 & 1.8 \\ 
c8f1n1 & 7.5 & 2.5 & 4 & 0.7 & 3.3 & 103.6 & 0.00 & 0.00 & 1.6 & 1.4 & 3 & 0.3 & 0.4 & 8.4 & 267.0 & 0.00 & 0.00 & 2.3 & 2.3 \\ 
c8f1n2 & 7.5 & 3.8 & 6 & 0.7 & 3.4 & 119.9 & 0.00 & 0.00 & 1.6 & 2.1 & 5 & 0.4 & 1.2 & 8.2 & 256.0 & 0.00 & 0.00 & 2.2 & 1.8 \\ 
c8f1n3 & 7.5 & 5.1 & 8 & 0.7 & 3.4 & 181.1 & 0.00 & 0.00 & 1.6 & 2.8 & 7 & 0.4 & 1.0 & 8.2 & 275.1 & 0.00 & 0.00 & 2.2 & 1.9 \\ 
c8f1n4 & 7.5 & 6.4 & 10 & 0.7 & 3.4 & 220.4 & 0.00 & 0.00 & 2.1 & 3.5 & 9 & 0.5 & 0.7 & 8.1 & 179.4 & 0.00 & 0.00 & 2.2 & 2.1 \\ 
c8f2n1 & 7.5 & 2.6 & 4 & 0.7 & 3.3 & 103.6 & 0.05 & 0.05 & 1.6 & 1.4 & 3 & 0.7 & 0.9 & 8.9 & 27.0 & 0.05 & 0.11 & 1.9 & 2.0 \\ 
c8f2n2 & 7.5 & 3.9 & 6 & 0.7 & 3.4 & 130.7 & 0.05 & 0.05 & 1.6 & 2.1 & 5 & 0.7 & 1.4 & 8.5 & 51.9 & 0.05 & 0.11 & 2.0 & 1.8 \\ 
c8f2n3 & 7.5 & 5.3 & 8 & 0.7 & 3.4 & 195.8 & 0.05 & 0.05 & 1.6 & 2.8 & 7 & 0.8 & 1.2 & 8.3 & 44.3 & 0.05 & 0.10 & 1.9 & 1.8 \\ 
c8f2n4 & 7.5 & 6.6 & 10 & 0.7 & 3.4 & 225.8 & 0.05 & 0.05 & 1.7 & 3.6 & 9 & 0.8 & 1.1 & 8.2 & 59.5 & 0.05 & 0.10 & 2.0 & 1.9 \\ 
c8f3n1 & 7.5 & 2.7 & 4 & 0.7 & 3.3 & 103.0 & 0.10 & 0.10 & 1.6 & 1.4 & 3 & 0.9 & 1.3 & 8.9 & 18.4 & 0.09 & 0.20 & 1.8 & 1.8 \\ 
c8f3n2 & 7.5 & 4.0 & 6 & 0.7 & 3.3 & 133.1 & 0.10 & 0.10 & 2.1 & 2.2 & 5 & 0.8 & 1.4 & 8.6 & 33.2 & 0.09 & 0.19 & 1.9 & 1.8 \\ 
c8f3n3 & 7.5 & 5.4 & 8 & 0.7 & 3.4 & 199.3 & 0.10 & 0.10 & 1.6 & 2.9 & 7 & 0.9 & 1.3 & 8.4 & 40.9 & 0.09 & 0.18 & 1.9 & 1.8 \\ 
c8f3n4 & 7.5 & 6.8 & 10 & 0.7 & 3.4 & 230.1 & 0.10 & 0.10 & 1.7 & 3.7 & 9 & 0.9 & 1.1 & 8.2 & 42.5 & 0.09 & 0.17 & 1.9 & 1.9 \\
\enddata
\end{deluxetable}

We thank Jarrod Hurley for help with the BSE code and 
Paolo Miocchi for providing us with his fitting codes for single-mass 
King models.  
This work was supported by NASA Grants
NNX08AG66G and NNG06GI62G at Northwestern University.  
JMF acknowledges support from Chandra/Einstein Postdoctoral Fellowship Award 
PF7- 80047.
This research
was partly done at KITP while the authors participated in the Spring 2009 program on
ÒFormation and Evolution of Globular ClustersÓ, and was supported in part by NSF
Grant PHY05-51164.
      
\bibliography{biblio_paper5}

\end{document}